\documentstyle[12pt,astron,psfig,huub]{article_huub}

\begin{document}

\centerline{\Large\bf Spectroscopy of Ultra--Steep Spectrum Radio Sources:} 
\centerline{\Large\bf A Sample of $z>2$ Radio Galaxies}

\begin{verse}
H.J.A. R\"ottgering$^{1,2,3}$, R. van Ojik$^1$,
G.K. Miley$^1$, K.C. Chambers$^4$,
W.J.M. van Breugel$^5$, S. de Koff$^{1,6}$
\end{verse}

{\small\it
\begin{quote}
\item $^1$Leiden Observatory, P.O. Box 9513, 2300 RA, Leiden, The Netherlands 
\item $^2$Mullard Radio Astronomy Observatory, Cavendish Laboratory, 
Madingley Road, Cambridge, CB3 0HE, England 
\item $^3$Institute of Astronomy, Madingley Road, Cambridge, CB3 0HA, England 
\item $^4$Institute for Astronomy, University of Hawaii, 2680 Woodlawn 
Drive, Honolulu, Hawaii 96822, USA
\item $^5$Institute for Geophysics and Planetary Physics,
Lawrence Livermore National Laboratory, L-413, Livermore, CA 94550, USA
\item $^6$Space Telescope Science Institute, 3700 San Martin Drive, Baltimore MD21218, USA
\end{quote}
}

\vspace{6cm}

\noindent H. R\"ottgering et al.: Spectroscopy of Ultra--Steep Spectrum Radio Sources

\noindent Astronomy and Astrophysics, Main Journal

\noindent 3. Extragalactic Astronomy

\noindent 11.01.2, 11.01.4

\bigskip 

\noindent H. R\"ottgering, 

\noindent  Leiden Observatory, P.O. Box 9513, 2300 RA, Leiden, The Netherlands 

\newpage
\addtocounter{page}{-1}
%\thispagestyle{empty}
%\centerline{\Large\bf Spectroscopy  of Ultra--Steep Spectrum Radio Sources } 
%\centerline{\Large\bf A sample of $z>2$ radio galaxies}

\bigskip

\begin{abstract}
\noindent
We present spectroscopic observations for 64
radio galaxies having ultra steep radio spectra.
Twenty-nine objects have redshifts $z>2$, the largest redshifts being almost 4.
Our ultra steep spectrum (USS) criterion ($\alpha < -1$) has proven to be the
most efficient way of finding distant radio galaxies.
We find that even among the USS sources,
there is a strong statistical correlation between the spectral index 
and redshift. The most distant radio galaxies within the USS sample 
have the steepest radio spectra.

In our sample there are 3 radio galaxies at $z>3$ compared with 26 at $2 < z < 3$.
However, the present data do not allow us to decide 
whether there is a decrease in co-moving source density at the highest redshifts.

We have analyzed the spectra of the 30 objects with the highest redshifts
($z>1.9$). For these high redshift radio galaxies, Ly$\alpha$ is almost always the
dominant emission line, with a rest frame equivalent width ranging from
$\sim100$ \AA\ to more than 1000 \AA. The equivalent widths of the most
important emission lines (Ly$\alpha$, C\,{\small IV}, He\,{\small II}, 
C\,{\small III}]) are found to correlate
strongly with each other. The large rest frame equivalent widths and the
correlation between the equivalent widths of the emission lines, confirm that
photoionization by a central continuum source is most likely the dominant
ionization mechanism.

There are significant velocity differences between the 
various emission
lines of our high redshift radio galaxies; 
in particular the Ly$\alpha$ line is shifted with
respect to the higher ionization lines. Velocity shifts range from 100 to
almost 1000 km s$^{-1}$ in some cases.
Simulations show that the effects of associated H\,{\small I} absorption on the
Ly$\alpha$ emission line may be responsible
for most of these velocity shifts. However,
other mechanisms such as organized kinematics of
the Ly$\alpha$ emission line gas (e.g. inflow or outflow) and obscuration
of the line emission from the far side of the radio galaxy may also  play a role.
\end{abstract}


\section{Introduction}
The most distant observable galaxies are radio
galaxies. Because these objects can be seen to large redshifts and
because they are spatially extended, they are unique probes of the
early Universe and important laboratories for testing models of galaxy
formation.
The most efficient way of obtaining significant samples of
radio galaxies at redshifts $z>2$ is to concentrate on the
counterparts of radio sources with ultra-steep spectra (USS, $S \sim
\nu^\alpha, \> \alpha \lta -1.0$; e.g. Miley \& Chambers 1989;
\nocite{mil89} Chambers \& Miley 1989).  \nocite{cha89b}  During the last
5 years more than 60 radio galaxies with $z>2$ have been discovered
(including those described in this paper), mostly
using this technique. The most distant 
USS radio galaxy discovered so far is 8C1435+633 at $z=4.25$ 
\cite{lac94}.

High redshift radio galaxies (HZRGs) exhibit a large variety of
properties. Remarkable is the initially unexpected alignment between 
the UV/optical emission and the radio structures and the common presence
of enormous ($>100$ kpc) ionized gas halos for 
which the  Lyman $\alpha$ emission line is very strong 
(e.g. McCarthy et al. 1993).
The variety of emission components from HZRGs and their large luminosities
provide unique diagnostics for studying galaxies in the early Universe.

In order to study the nature of HZRGs and their evolution, it is
important to obtain a large sample of such objects. After the initial 
success of the high redshift search technique on a sample of 4C USS sources
(8 of 33 had $z>2$, Miley \& Chambers 1989;
\nocite{mil89} Chambers \& Miley 1989\nocite{cha89b}), 
we set about increasing the number of HZRGs by extending
the USS search method to larger and fainter samples of radio sources.
The resultant Leiden intermediate-flux compendium of USS sources was the
basis for a key programme carried out at the European Southern Observatory in 
Chile between 1990 and 1994. This programme was remarkably successful, resulting
in the discovery of 22 new radio galaxies with $z>1.9$.

The project consisted of several steps.  First, samples of
USS sources were compiled having 
a range of   finding frequencies and 
flux densities (see R\"ottgering et al. 1994\nocite{rot94}.
Secondly, high resolution ($1.5 ''$)
imaging of USS sources with the VLA was carried out (see R\"ottgering et al. 
1994)\nocite{rot94} to obtain good
positions so that the USS sources could be identified on deep CCD images.  
Thirdly, a subset of the sources
was imaged in $R$ band with 2-m class telescopes to obtain reliable
identifications (see R\"ottgering et al. 1996a).\nocite{rot95})
Finally, low-resolution spectroscopy was carried out with 4m-class
telescopes on a subset of the objects to
determine their redshifts. Here we present the results of this spectroscopy.
We have augmented it with previously unpublished data from the original 4C 
sample in order to carry out some initial statistical analyses.
The spectra presented here have been the starting point for several 
follow-up programmes to investigate various properties of HZRGs, some of which 
are discussed elsewhere in this thesis.

The structure of this paper is as follows.  In Section 2, we describe
the sample selection and in Section 3 the observations and reduction
of the data.  The
redshifts and spectra are presented in Section 4. We also
analyze the properties of the emission lines in this section, 
determining velocity shifts between the emission
lines and searching for correlations with other
properties of the radio galaxies. In Section 5 we discuss our results.
We consider the
efficiency of the USS technique for finding high redshift radio
galaxies and briefly discuss the radio luminosity function of USS sources
at high redshifts.
Furthermore, we investigate a possible origin for the observed 
velocity shifts of Ly$\alpha$ with respect to the higher ionization lines
as being due to associated H\,{\small I} absorption systems.
In Section 6 we summarize our results and conclusions.

Throughout this paper we assume a Hubble constant of
$H_0=50$ km s$^{-1}$ Mpc$^{-1}$ and a deceleration parameter of $q_0=0.5$.

\section{Samples}

The sources whose optical spectroscopy we shall discuss here, have
been selected either from the compendium of USS sources as compiled by
R\"ottgering et al. (1994)  or from the 4C sample of USS sources as
presented by Tielens et al. (1979). \nocite{tie79}

\subsection{Leiden USS compendium} 

The Leiden compendium of USS radio sources (R\"ottgering et al.  
1994\nocite{rot94})
contains sources from
a variety of radio catalogues, including the 8C at 38 MHz 
\cite{ree90b}, the 6C at 151 MHz \cite{hal88a}, the
Texas Sky Survey at 365 MHz \cite{dou80}, the Molonglo
Catalogue at 408 MHz \cite{lar81}, the 1400 MHz NRAO Sky
Survey Condon \cite{con85,con86} and the 4.85 GHz Sky 
Survey \cite{con89a}.  The resulting compendium therefore
contains sources with a range of finding frequencies.  The median flux
density at 365 MHz is $\sim 1$ Jy.
A large fraction of the sources from this compendium has been imaged
at $R$-band to a level of $R > 24$ using 2-m class telescopes at ESO
and La Palma (e.g. R\"ottgering et al.  1996a) \nocite{rot95}

Since our main aim was to obtain a sample of $z>2$ radio galaxies for
subsequent detailed studies, we selected sources for
spectroscopy so as to optimize the chance of finding high redshift
objects. In carrying out the project, compromises had to be made in the
selection of targets. It was regarded as of paramount importance to increase
the {\it number} of high redshift radio galaxies. In order to optimize the 
use of the scarce 4m-class telescope time towards this goal, additional 
selection criteria were applied to reduce the number of spectroscopic targets.
While recognizing that these additional criteria introduce complications 
on subsequent statistical analyses, we regarded it as more important at this 
stage to increase the number of high redshift galaxies whose properties could
be studied in detail. Two additional radio-based criteria were therefore used 
in optimizing the spectroscopic targets for high redshift objects.
First, we excluded the counterparts of diffuse radio sources. Diffuse
(Fanaroff-Riley Class 1, Fanaroff \& Riley 1974)\nocite{fan74a} radio 
morphologies are known to be associated with relatively weak (nearby) radio 
sources. Secondly, we concentrated on the counterparts of smaller ($<20''$)
radio sources. Taking account of the well-known correlation of radio angular 
size with redshift for quasars \cite{bar88b}, only 25\% of $z>2$ steep 
spectrum quasars have angular sizes $>10''$. Hence, even if radio galaxies
are systematically larger by a factor of 2, this criterion should exclude
relatively few $z>2$ objects. Finally, a selection on optical magnitude from
the CCD imaging was applied, concentrating on the objects with faint
optical identifications ($R>21$) as the best candidates for distant 
radio galaxies.

\subsubsection{4C USS sample} 
 
The 4C USS sample comprised 33 sources from Tielens et al. (1979) that have
spectral indices $\alpha < -1$ between 178 and 5000 MHz. Further details on
the sample selection and observations can be found in Chambers \& Miley (1989);
\nocite{cha89b} Chambers, Miley and van Breugel (1987) and
Chambers et al (1996a). \nocite{cha87,cha96a} 
Within this sample 8 have $z>2$ \cite{cha96b}, including 4C40.36 at $z=2.3$
\cite{cha88a}, 4C41.17 \cite{cha90} and 4C48.48 at $z=2.3$ \cite{cha89b}. Here
we will present the spectral properties for these 8 sources. Their
median flux density at 365 MHz is $\sim 2$ Jy.

\section{Observations of the Leiden USS sample}
The observations were carried out with CCD slit spectrographs on 
the 3.6m telescope and the NTT (3.5m)
of the European Southern Observatory (ESO) at La Silla
and on the WHT (4.2m) of the Observatorio del Roque de los Muchachos
at La Palma.
The ESO observations were carried out as part of a ``Key Programme''.
Details of the observational setup  are given in Table 1.
Typical observing times were  1--2 hours per object.

The emission line regions of distant radio galaxies are
typically extended by at least several arc seconds and oriented along the
radio axis (e.g. McCarthy et al. 1990b\nocite{mcc90b}). In order to optimize the
signal to noise for detecting such emission lines, we therefore
usually aligned the slit in the direction of the radio
source and used a relatively large slit width ($> 2''$). 
The typical emission line widths of distant radio galaxies is $\sim
1000 - 1500$  km s$^{-1}$. 
With the slit width and the gratings used (see
Table 1), the resolution of the spectra are comparable to
the expected width of the emission lines. 

Since the targets were too faint to be seen on the TV guiders, the
following procedure was used for positioning the objects in the slit.
The slit was first centered on either a reference star from
the Space Telescope Guide Star System (GSS) or from a star on the
R-band CCD frames \cite{rot95},
whose position had been accurately
determined with respect to the GSS system. 
A short exposure was made
to check that this fiduciary star was indeed centered in the slit and
the telescope was then offset to the target.  Since the relative
position between the target and the positional reference star is known
with an uncertainty of less than $0.5''$, the target was assumed to be
in the slit.

\subsection{Reduction and Analysis}

Reduction of the spectra was carried out using the `Long-slit' package
in the IRAF image reduction package developed by the US National
Optical Astronomy Observatory (NOAO).  For the data taken on every
night, first a bias was constructed averaging 10 ``zero second''
exposures taken at either the beginning or at the end of each
night. This bias was subtracted from every non-bias frame.  The
pixel-to-pixel gains were calibrated using flat fields obtained from
an internal quartz lamp.  Wavelength calibration was carried out by
measuring the positions on the CCD of known lines from either an He-Ne
or a Cu-Ar calibration lamp, fitting a polynomial function to these
data, and applying the resultant calibration factors.
We estimate the accuracy of the wavelength calibration 
to be about 5 \% of the resolution, i.e. $\sim 1$ \AA\ 
(see Table 1).

The sky contribution was removed from the frames, by subtracting a sky
spectrum obtained by fitting a polynomial to the intensities measured
along the spatial direction on the two-dimensional spectrum not including
the spatial rows/columns where the targets were positioned.
One-dimensional spectra were extracted
by averaging in the spatial direction over an effective aperture as
large as the spatial extent of the brightest emission line.
Each night
one or two standard stars were observed and used for the
flux-calibration. We estimate that the flux-calibration is good to $\sim
10\%$.

\subsection{Identification procedure} 

After all the geometrical and intensity calibrations were applied and the
sky background level was subtracted, identification of the various emission
lines in the spectra was attempted.

Depending on their redshifts, the emission line spectra of radio
galaxies  are dominated by one or several of the following lines:
Ly$\alpha$ $\lambda$  1216, [O\,{\small II}]
$\lambda$ 3727, [O\,{\small III}] $\lambda$5007, 
H$\alpha$ $\lambda$ 6563 + [N\,{\small II}]
$\lambda$ 6583
\cite{spi86,ost89,mcc93a}.
For the more distant ($z>1$) radio galaxies other lines that are
characteristic of the spectra, but with fainter intensities (5 - 10\%
of Ly$\alpha$)
are C\,{\small IV}, He\,{\small II}, C\,{\small III}], C\,{\small II}], 
[Ne\,{\small IV}], Mg\,{\small II} and [Ne\,{\small V}].

The procedure for determining the redshifts consisted of first
attempting to identify detected emission lines with the dominant lines
and thereafter searching for
fainter lines at the appropriate redshift.

An additional consideration is the presence of the Ly$\alpha$ ``break'' in the
continuum spectrum due to the Ly$\alpha$ forest or a decrease in the intensity
of the stellar continuum bluewards of Ly$\alpha$.  The presence of such 
a break at the appropriate wavelength is an additional argument
for identifying a line with Ly$\alpha$.

We note that 1243+036 (z=3.6) is the 
only object in the $z>2$ sample for which we have 
detected only one emission line. For a number of reasons we 
are confident that the redshift is correct, the most important 
being that [OIII] 5007 emission is clearly detected in the 
infrared at the expected wavelength 
(cf van Ojik et al 1996a). 

\subsection{Analysis of the spectra} 
After an initial identification of the lines, each spectrum was
analyzed and relevant parameters of the lines were determined. 
The first step was fitting the line with a Gaussian profile superimposed
on a constant continuum.
From this, the peak intensity and the width (FWHM) of the line were
obtained.

The effect of noise on each spectrum was determined by calculating the RMS intensity
in a line-free spectrum, i.e. excluding wavelengths within twice the 
measured FWHM of the emission lines or residual sky lines.  
The RMS intensities in these resulting spectra were 
calculated in bins of about 25 pixels. The values for the
rms in each bin were fitted with 4th order polynomials. These polynomials
then give an estimate of the local rms at each location in the spectra.

The uncertainty in the wavelength of the intensity peak of an emission line
is a combination of the systematic uncertainties in the
wavelength calibration, estimated to be 5 \% of the resolution, and
the uncertainty in the measurement of the peak position, estimated to be
(FWHM $/ (2 s/n) $), where $s/n$ is the signal to noise ratio of the
observed emission line (e.g. Condon 1989).\nocite{con89}  
The weighted average of the redshifts of the peaks of the
emission lines has been taken as the redshift of the radio galaxies.  
The uncertainties in the redshift determination were calculated in two ways.
First, it was calculated from a weighted average of  the
uncertainties in the peak positions of the identified 
lines. Secondly,  it was calculated from the differences between the 
observed 
redshift of the source and the redshifts of the individual 
lines.  Since some of
the lines have velocity shifts 
with respect to other lines (see below), these two
errors are not always in agreement. The uncertainty in the redshift, given 
in Table 2, is the largest of these two values.
Especially the Ly$\alpha$ emission line may be affected by HI absorption
so that the true centroid of the Ly$\alpha$ emission line is not well determined
by a Gaussian fit. This may cause a different redshift for the Ly$\alpha$ 
line than for the other emission lines in the spectrum 
(see Sects. 4.2 and 5.4),
thus contributing to the uncertainty in the redshift.

The deconvolved widths (FWHM) 
of the lines were determined assuming that the lines
are Gaussian.  We note that from observations of the Ly$\alpha$ emission at
resolutions of a factor 10 higher than the observations presented here
it is clear that the line profiles can be very complex \cite{oji96a,oji96b}.
The deconvolved width presented here is therefore only a rough
estimate of the true width.  The formal uncertainty 
in the deconvolved width (FWHM) is
a combination of (i) the uncertainty due to the signal to noise ratio of 
the line, estimated to be
(FWHM$ / (s/n) $) (e.g. Condon 1989) and (ii) the uncertainty in the
resolution, estimated to be 10 \%.
If the fitted FWHM is less than the resolution or the determination of
the deconvolved FWHM is less than its error, then the error in the
deconvolved FWHM is taken as the upper-limit to the deconvolved FWHM. 

The line flux was obtained by summing the intensities of the pixels
above the determined continuum over a wavelength range 
of size 4 times the width (FWHM) of the
line and centred at the peak of the emission line.  The error in the
flux determination is calculated taking into account (i) the signal to
noise ratio of the line and (ii) the error in the flux calibration,
estimated to be 10\%.  We note that in general the emission region
will be significantly more extended than the width of the slit and
that therefore the fluxes of the lines are a lower limit to the total
flux of the emission line.

The rest-frame equivalent width of each line was determined as 
the ratio of the line
intensity to the local continuum intensity divided by $(1+z)$. 
Since the continuum emission from
our targets is faint, 
in a significant number of cases only an upper limit to the continuum 
could be determined, yielding a lower limit to the equivalent width.

In general the spatial extent of the Ly$\alpha$ emission line was
observed to be 
significantly larger than the point spread function determined by the 
seeing. The angular size was determined by 
measuring the distance (in arcseconds) between the most extreme spatial points
where Ly$\alpha$ was detected on the two-dimensional spectrum, i.e. at a limiting
surface brightness in the spectra of typically 10$^{-18}$ erg s$^{-1}$ cm$^{-2}$
\AA $^{-1}$ arcsec$^{-1}$.
This measurement is clearly dependent on parameters such as seeing and signal
to noise ratio of the line and therefore the tabulated values 
should be taken only as an indication
of the size.  Furthermore, it is clear from narrow band imaging that the
morphology of the Ly$\alpha$ emission is usually complex and non-spherical.
Hence, extensions in directions other than
the slit direction (along the radio axis) will differ from the tabulated values
presented here.
The higher ionization lines were usually not detected with enough signal 
to noise to decide whether they were spatially resolved. They may be similarly
extended as the optical continuum emission from the radio galaxies (usually 2--3$''$).

\section{Results}

\subsection{Redshifts}
From our spectroscopic observations we have found 
redshifts for a total of 64 USS radio sources, of which
29 are at redshifts larger than 2.
In Fig. 1 the individual spectra of the objects with $z>1.9$ are 
presented. 
The lines that we have identified are indicated in the plots of the spectra.
We also have indicated with the symbol $\odot$, where 
the spectra are severely affected by badly subtracted sky-lines. 
In a large fraction of such cases we excluded the 
affected pixels and therefore the residual sky-lines 
are absent on the plots of the spectra. 
In Table 2 the redshifts of the objects with $z \geq 1.9$ plus the 
relevant parameters of the emission lines in the spectra are presented.

Table 3 lists the sources with redshifts smaller than 1.9.
In the case of three objects for which only one emission 
line was detected, the redshift 
needs confirmation and is shown in parentheses.
The redshift distribution of our sample of radio galaxies is shown 
in Fig. 2.

From a few objects broad emission lines were observed ($> 4000$ km s$^{-1}$
FWHM) and they are therefore probably quasars. The broad line objects 
with $z>2$ 
are 1113$-$178 and 1436+157. The latter object is a quasar with a possible
companion galaxy within the extended Ly$\alpha$ halo of the quasar, 
oriented along the radio axis.
Because no obvious line emission was detected from the companion, we cannot 
definitely say whether this is an associated quasar--galaxy pair.

Statistical evidence for the preferential occurrence of companion galaxies
in the direction of the radio axis was provided by
the $R$ band images of our sample of radio sources \cite{rot95,rot95b}. However,
line emission was usually observed only from one of both objects, as in the
case of 0828+193 at $z=2.572$ and the above mentioned 1436+157 at $z=2.538$.
Therefore for these individual cases it is unclear whether both objects 
are at the same redshift. Only in 1707+105 at $z=2.349$ is the Ly$\alpha$ 
emission observed to have two clear
spatial peaks corresponding to the $R$ band positions of two galaxies oriented
along the radio axis.

\subsection{Emission line properties}

In the following we shall investigate the properties of the emission
lines and the presence of correlations between the
derived emission lines properties of the radio galaxies with redshifts
larger than 1.9 (all objects from Table 2).
We do not include the two objects with broad ($>4000$ km s$^{-1}$
FWHM) emission lines in the analysis. 
Thus, the sample for which we search for correlations of
the properties consists of 28 HZRGs.

\subsubsection{Ly$\alpha$ properties}
Ly$\alpha$ is the strongest and most spatially extended 
emission line in the spectra of our HZRGs. 
The measured spatial extent of Ly$\alpha$ along the slit ranges
from a few arcseconds to 16 arcseconds, corresponding to $\sim 25$ to 140 kpc.
In Fig. 3 we show the distribution of Ly$\alpha$ luminosities. 
The large
spatial extent of Ly$\alpha$ along the slit indicates that the Ly$\alpha$ extent
perpendicular to the slit may be larger than the slit width ($\sim 2.5''$) .
Therefore, it is
possible that the true Ly$\alpha$ luminosities are a factor of $\sim 2$ higher
than measured in our spectra.

Roughly 80\% of the radio galaxies have Ly$\alpha$
luminosities larger than 10$^{43.5}$ erg s$^{-1}$. Given the sensitivities
of our spectroscopy (see R\"ottgering 1993), this is unlikely to be 
a selection effect. 
The Ly$\alpha$ luminosities are comparable to the radio
luminosities of the radio galaxies and also comparable to the
luminosities of extended Ly$\alpha$ emission observed in high redshift
radio loud quasars
\cite{hec91b}.

Assuming a volume filling factor of the emission line gas of 10$^{-5}$ as 
derived from sulphur emission lines of low redshift radio galaxies 
\cite{bre85d,hec82},
the Ly$\alpha$ luminosities of the radio galaxies imply masses in ionized 
gas of a few times 10$^7$ to a few times 10$^8$ M$_{\odot}$ (see also 
van Ojik 1995a and 1995b).

\subsubsection{Equivalent widths}
The equivalent widths of the various emission lines (Ly$\alpha$, 
C\,{\small IV}, He\,{\small II}, C\,{\small III}])
are plotted against each other in Fig. 4.
In the cases where the continuum level could not be measured, lower limits
to the equivalent widths were determined. In a few cases where an emission 
line was not detected but the continuum could be measured, an upper limit 
to the equivalent width of the line was determined.

We find that the rest frame equivalent widths of all 4 lines are strongly 
correlated with each other. 
Spearman rank analysis of the correlations give confidence levels
of more than 99\%.
The strongest and most obvious 
correlations are visible between the equivalent widths of the higher ionization
lines (C\,{\small IV}, He\,{\small II}, C\,{\small III}]). 
The correlations with the Ly$\alpha$ equivalent
width are less  significant.

We note that most of the lower limits imply that the true correlations
between the equivalent widths might be even stronger than seen in
Fig. 4.  We found no evidence for correlations of the equivalent
widths with other properties of the radio galaxies within our sample
(radio size, luminosity, distortion, spectral index).

\subsubsection{Velocity shifts between emission lines}
In many cases we find that the redshifts of the peak intensities determined
using Gaussian fits to
the individual emission lines in a spectrum differ by more than the
combined errors in the redshifts of the individual lines. The rms-uncertainties
in
the wavelength calibration are typically less than 2 \AA, corresponding to
$\sim 130$ km s$^{-1}$, while the measured velocity shifts between the lines
are between several hundreds and sometimes almost $1000$ km s$^{-1}$.

To establish the reality of these shifts we must ensure that there are no
systematic errors in our wavelength determination.
An uncertain wavelength determination may occur when an 
emission line occurs
at a wavelength outside the range of the lines emitted by 
the calibration lamp, so 
that the wavelengths are determined by extrapolation rather than 
by interpolation from the 
calibration lines. This is the case for
0448+091, where Ly$\alpha$ is at 3692 \AA\ 
and the detections of C\,{\small IV}, He\,{\small II} and C\,{\small III}] are
marginal. In the observing
session during which 0448+091 was observed (WHT January 1991)
there were few calibration lines in the far blue. The bluest
usable calibration line was at 4131 \AA. Thus, the Ly$\alpha$ emission line
was 340 \AA\ blueward of the nearest calibration line. This extrapolation may
well have caused an erroneous determination of the wavelength of Ly$\alpha$
and we therefore ignore this galaxy in considering the
velocity correlations. During the other observing sessions there were
no such problems and calibration lines were available down to 3888 \AA.

Finally, there is the problem is the presence of strong skylines. In
the objects 0529$-$549, 0828+193 and 4C\,26.38 the C\,{\small IV}
emission line is affected by the 5577 \AA\ skyline. We have therefore
excluded these three objects in the following analysis of the velocity
shifts.

After also leaving out 1243+036 for which Ly$\alpha$ was the only
detected emission line in the spectrum, the final sample for which we have
carried out the velocity shift analysis consists of 23 HZRGs.
The determined velocity shifts between the various emission lines
and their uncertaintes are listed in Table 4.
In this table negative values indicate that the first line is redshifted
with respect to the second.
Although the uncertainties on most velocity shifts are large and the shifts
have only a 2--3 $\sigma$ significance, the presence of
velocity shifts in the sample is statistically significant (see below).

In Fig. 5 we show the velocity shifts between the emission lines
Ly$\alpha$, C\,{\small IV}, He\,{\small II} and C\,{\small III}] 
plotted against each other. 
Several correlations between the velocity differences are apparent:
strong correlations are present between the measured
velocity differences of the higher
ionization lines with respect to Ly$\alpha$. 
Objects with larger velocity shifts in one
high ionization line with respect to Ly$\alpha$ tend to show also larger
shifts of the other high ionization lines with respect to Ly$\alpha$. 
This is strong evidence that there is a real systematic shift between
the redshift of the peak intensity of Ly$\alpha$ with respect to the 
other lines.

From Fig. 5 it can also be seen that in almost all
objects C\,{\small IV} is redshifted with respect to both He\,{\small II} 
and C\,{\small III}].
The average shift of C\,{\small IV} with respect to He\,{\small II} 
is $-332\pm 101$ km s$^{-1}$
and C\,{\small IV} with respect to C\,{\small III}] 
is $437\pm 125$ km s$^{-1}$.
There may also be a correlation between the velocity differences of 
C\,{\small IV}
with He\,{\small II} and C\,{\small IV} with C\,{\small III}], 
although the errors on the individual points 
of the plot are larger than in the case of the first three correlations. 

One object that has a relatively large offset in the plots from the
other objects and relatively large uncertainties is 1138-262.  This 
radio galaxy is a very peculiar galaxy, both from its $R$
band image, where it is an extremely clumpy  
galaxy \cite{pen96}, and from a high
resolution radio image \cite{car95}, where it appears as a string of
radio knots. Furthermore, the high resolution Ly$\alpha$ spectrum
reveals that it has a very peculiar velocity structure
\cite{pen96}. Although it has these peculiarities which may be the
cause of its particular position in the velocity plots, we have no
reason to believe that our velocity determination has any systematic
error as in the case of 0448+091.

We find from a Spearman rank analysis that the correlations of velocity shifts 
of the higher ionization with respect to 
Ly$\alpha$ are significant at a confidence level 
of more than 99\%. The apparent 
(C\,{\small IV}--He\,{\small II})--(C\,{\small IV}--C\,{\small III}]) 
correlation would 
have a formal significance in a Spearman rank analysis of 95\%, but the 
errors on the individual points are too large to 
allow a Spearman rank analysis. 
Furthermore, 
if the previously mentioned radio galaxy 1138-262 is omitted, the correlation 
becomes insignificant (85\%). Thus, we conclude that this apparent correlation
is not statistically significant.
No significant correlations were found between the other velocity differences.
Thus, the three most significant correlations in the velocity shifts
indicate that there is frequently a velocity shift between the Ly$\alpha$ line
and the higher ionization lines.

Summarizing, we find that in many cases significant velocity shifts
between emission lines are present. Velocity shifts of Ly$\alpha$ with
higher ionization lines are strongly correlated and C\,{\small IV}
appears to be mostly redshifted with respect to He\,{\small II} and
C\,{\small III}].  A comparison with velocity shifts earlier reported
in quasars and radio galaxies
\cite{gas82b,wil84c,esp89,cor90,car91a,mcc90a,tad91,eal93b} and their
possible interpretation will be discussed in a next section.

\section{Discussion}

\subsection{USS selection efficiency}

The technique of selecting ultra steep spectrum radio sources,
has proven to be very successful in finding distant radio galaxies. 
In total we have obtained spectra of
147 objects, of which for 64 (44\%) we have been able to determine a redshift.
From the remaining objects no clear emission lines were observed. 
There are four possible explanations for the failure to detect emission lines in
these objects. First, the objects did not have emission lines of sufficient 
strength to be detectable in the given atmospheric conditions and sensitivities.
Secondly, many of these sources may be at redshifts between
$\sim 1.2$ and $\sim 1.6$, where there are no extremely 
strong emission lines visible
in the optical window. Thirdly, some objects may be at $z>5$, where Ly$\alpha$
is shifted beyond the red of the window to be observable.
Fourthly, there may have been 
some unseen errors in the telescope pointing.

Of the 147 objects observed, 29 (19.7\%) are at redshifts larger than 2, which
shows that the USS selection is indeed an efficient way of finding very distant
radio galaxies. Taking into account the preselections from VLA
imaging of $\sim 600$ sources and from $R$ band CCD imaging of $\sim 300$ objects,
we can say that of the radio sources with spectral index
$\alpha < -1$, radio size $<20''$ and faint optical identification ($R > 21$),
roughly 1 in 5 objects is at a redshift larger than 2.

To further assess the efficiency of the USS selection technique we examined
the relation between redshift and spectral index.
In Fig. 6 we plot this relation for a total
of 108 radio galaxies, where the spectral index is that defined between
the lowest available frequency above 150 MHz and 5 GHz. 
Apart from our USS sources (selected $\alpha < -1$), 
this plot also includes radio galaxies from the Molonglo Survey 
($\alpha < -0.9$, McCarthy {et al.} 1990a, 1990b\nocite{mcc90b,mcc90b}), 
the Bologna Radio catalogue ($\alpha < -0.9$, McCarthy {et al.} 1991;
Lilly 1988\nocite{mcc91d,lil88}), 8C \cite{lac94,lac92} and 6C \cite{eal93c}.

From this plot we see that there is a clear trend for the higher redshift 
sources to have steeper radio spectra. The median spectral index of all sources
is $-1.15$. At $z<1$ the median spectral index is $-1.08$, at redshifts between
1 and 2 it is $-1.16$, at redshifts between 2 
and 3 it is $-1.19$ and at $z>3$ the median spectral index is $-1.31$
A Spearman Rank correlation analysis shows that the correlation between
spectral index and redshift is present at a confidence level of more than
99.9\%. If only considering the radio galaxies at redshifts larger than 1,
this correlation is still significant at the 98\% confidence level.

Particularly, for the steepest spectra the efficiency of the technique in 
finding the most radio distant radio galaxies is remarkable.
Of all sources with radio spectra steeper than $-1.3$ 
almost 30\%$\pm$15\% are at redshifts
larger than 3, compared with only 3\%$\pm$2\% 
of the sources with spectral index between $-0.9$ and $-1.3$.

This continued steepening of the observed radio spectra with redshift
can be understood as the effects of the ``K--correction'' at radio
wavelengths: There is a well known
tendency of radio sources to have spectra which steepen at higher
frequencies. For objects at higher redshifts the receivers sample
higher emitted frequencies and steeper spectra.
Evidence that the intrinsic radio spectra of HZRGs indeed continue to steepen 
to higher frequencies comes from 5 and 8 GHz VLA observations of our sample 
\cite{car95}. An additional effect that could contribute is 
that the radio emission would
have an intrinsically steeper spectrum at a fixed rest frame frequency due
to increased inverse Compton losses from scattering of the microwave background,
which is stronger at higher redshifts (Krolik \& Chen 1991; 
see also Blumenthal \& Miley 1979; Chambers et al. 1990).
\nocite{kro91,blu79,cha90}

Further it has been suggested that the spectral index might be linked to the
total radio power \cite{lai80a} possibly due to the density of the environment.
Because we are observing a flux limited radio sample, the objects at the
highest redshifts may have a larger average radio power. However, at $z>2$
the range in radio power of the sources is relatively small.
There is no evidence for a correlation between radio power and the spectral
index of the $z>2$ sources (see Carilli {et al.} 1996)\nocite{car95}. 
Although a mechanism in which a denser environment of a radio source (as might be
expected around primeval galaxies in the early Universe) may cause the radio
lobes to have a steeper spectrum, the
most plausible cause for the strong relation of spectral index with redshift,
remains the steepening of the radio spectrum at higher rest frame frequencies
with a possible contribution from 
increased inverse Compton losses due to the more intense microwave
background at high redshifts.

\subsection{The evolution of the space density of radio galaxies}

Although there appears to be an abrupt decrease in the observed number
of radio galaxies at redshifts $z>3$, this does not necessarily imply
a real decrease in the space density of radio galaxies at those
redshifts (e.g. Dunlop \& Peacock 1990). \nocite{dun90} A nice way to
test for evolution beyond a defined redshift $z_0$ is to use a modified
version of the $<$$V/V_{max}$$>$ test, the ``binned''
$<$$V'/V'_{max}$$>$ (e.g. Osmer \& Smith 1980)\nocite{osm80}.

For such a test it is important that the flux limit of a survey is
well established.  The flux limits for our samples of USS sources are
not well defined for three reasons.  First, our sample from the Texas
Catalogue and the sample from the 4C survey do not have a well
determined flux limit. At low flux levels ($<0.5$ Jy for the Texas
Catalogue and $<2$ Jy for 4C) the surveys are incomplete.  Secondly,
the actual flux limit for the Texas Catalogue is a function of source
morphology. Thirdly, the selection on spectral index by combining two
radio catalogues increases the flux limit at low frequencies for
sources with steeper spectra.  

This poorly known flux limit introduces an error in the
$<$$V'/V'_{max}$$>$ of at least 0.1--0.2 (see R\"ottgering
1993)\nocite{rot93}.  Therefore we conclude that our present sample of
USS radio galaxies, does not allow us to decide whether there is a
decrease in source density at the highest redshifts.

\subsection{Ionization of the emission line gas}
Three important ionization mechanisms that have been considered for
the emission line regions of radio galaxies are shock ionization from
the interaction with the radio lobes, photoionization by hot stars and
photoionization by the active nucleus.  In general, the line ratios of
radio galaxies and the rest frame equivalent width of
Ly$\alpha$ are not well reproduced by shock ionization models,
nor by hot stars, and photoionization by a nuclear UV continuum
appears to be the dominant mechanism of ionization
\cite{bau89b,bau89c,bau92,mcc93a,fer86,fer87,cha90,chr93}.

The strong correlations between the equivalent width of the emission lines
found for our sample 
is further support that the emission lines are all produced by the same
ionization mechanism and are consistent with central photoionization.

\subsection{Velocity shifts: possible associated absorption systems}

Velocity shifts between different emission lines in the spectrum of
quasars occur frequently.  For the broad lines of quasars, the high
ionization lines and Ly$\alpha$ are systematically blueshifted with
respect to the low ionization lines and H$\alpha$ by as much as 4000
km s$^{-1}$ \cite{gas82b,wil84c,esp89,cor90,car91a}.  These velocity
shifts have been attributed to outflow of highly ionized gas from the
active nucleus in combination with obscuration of the far side of the
quasar's emission line region.

The characteristics of the velocity shifts in our HZRGs are quite
different from those of quasars. Not only are the shifts generally
smaller, but also the shifting of the peak of the Ly$\alpha$ line with
respect to the peak of the higher ionization lines is not observed in
quasars, where Ly$\alpha$ together with the higher ionization lines
are shifted with respect tot the lower ionization lines and
H$\alpha$. Furthermore, in quasars the shifts are observed in the
broad line region, i.e. on a very different scale than the narrow line
region of radio galaxies.  It is therefore plausible that the origin
of the velocity shifts between the emission lines in HZRGs is
different from that in quasars.

We consider two explanations for the occurrence of the velocity shifts
in our sample of HZRGs. 
First, the spatial distribution and the velocity field of the high
ionization lines and the Ly$\alpha$ emission lines are not necessarily
the same and could lead to the observed 
velocity shifts.  In radio galaxies there have been reports of velocity
shifts of up to $\sim 1000$ km s$^{-1}$ between the emission lines
\cite{mcc90a,tad91,eal93b}. 
These have also been proposed to be related to the nuclear activity
\cite{tad91} or to be due to inflow or outflow on the near side of the
galaxy, with obscuration of the emission from the far side of the
galaxy \cite{eal93b}.

Secondly, a large fraction (60 \%) of HZRG show the presence of strong
associated H\,{\small I} absorption in spectra of Ly$\alpha$ emission
at a resolution a magnitude higher than the spectra presented in this
paper (R\"ottgering et al. 1995a; van Ojik et
al. 1996b). \nocite{rot95a,oji96b} In most cases the Ly$\alpha$
emission is absorbed at wavelengths blueward of the peak of the
Ly$\alpha$ emission, resulting in a red-shift of the peak of the
Ly$\alpha$ emission when observed at low resolution.

Since the observed correlations appear to be mainly due to the
Ly$\alpha$ emission line shifting with respect to the higher
ionization lines, this indicates that H\,{\small I} absorption systems
may play a role in producing the observed velocity shifts.
In the next paragraph this will be further considered.

\subsubsection{Simulations of Ly$\alpha$ velocity shifts caused by HI absorption}

Because of resonant scattering of Ly$\alpha$ photons, associated H\,{\small I}
absorption systems can produce a strong absorption feature in the Ly$\alpha$
emission profile that alters the shape of the line. When observed at low
spectral resolution 
this may result in a slightly different redshift for the Ly$\alpha$
line than for the higher ionization lines. An absorber located at a small
blueshift (or redshift) with respect to Ly$\alpha$ will cause a slightly
higher (or lower) redshift for the measured peak of Ly$\alpha$. To investigate
whether the effect of associated H\,{\small I} absorption systems on
Ly$\alpha$ could indeed produce the range of observed velocity shifts, we have
simulated the profile which would be observed if a Ly$\alpha$ emission line
with associated H\,{\small I} absorption was observed at low spectral
resolution.

We assume that the original emission line profile of Ly$\alpha$ is Gaussian.
An absorption feature is produced in the emission line profile 
by convolving it with a 
Voigt profile for a given H\,{\small I} column density and offset velocity from the
Ly$\alpha$ emission peak.
The resultant theoretical profile of Ly$\alpha$ emission with absorption feature
is then convolved with the instrumental resolution.
The peak position of this simulated ``observed'' profile is determined and gives
the offset velocity with respect to the peak of the original
Gaussian Ly$\alpha$ emission line profile.

We have carried out the simulations for an instrumental spectral 
resolution of 20\AA\ (FWHM) which is representative for our observations,
and an observing wavelength of 4000 \AA\ for Ly$\alpha$, 
so that the instrumental 
resolution is 1500 km s$^{-1}$.
The parameters that influence the resultant Ly$\alpha$ velocity shift are
the offset velocity of the H\,{\small I} absorption system, the H\,{\small I} column density,
the Doppler parameter ($b$) of the absorber, and the width of the original Gaussian
Ly$\alpha$ emission line.
From the high resolution observations of 0943$-$242 \cite{rot95a} it was
found that the H\,{\small I} absorption system has a column density of 
$\sim 10^{19}$ cm$^{-2}$
with a Doppler parameter of 55 km s$^{-1}$, while the velocity width of the 
original Ly$\alpha$ emission profile was estimated to be 1575 km s$^{-1}$. 
The velocity width of Ly$\alpha$
is usually in the range of 1000--1500 km s$^{-1}$ (FWHM) (e.g. McCarthy 1993; see also
Section 3 and van Ojik et al. 1995b)\nocite{mcc93a}.
Therefore, we shall assume similar parameters of the Ly$\alpha$ emission and 
H\,{\small I} absorption
systems for our simulations.

Column densities range from 10$^{14}$ to 10$^{21}$ cm$^{-2}$, Doppler
parameters of the absorption systems range from 25--75 km s$^{-1}$ and
the offset velocity of the absorber is varied from 0 to 1500 km s$^{-1}$
We have performed the simulations for input Gaussian emission lines of
widths 1800, 1500 and 1000 km s$^{-1}$ (FWHM).

Figure 7 shows the resultant velocity shifts for the observed
Ly$\alpha$ profile as a function of absorber offset.
The solid lines are for column densities 10$^{17}$, 10$^{19}$ and 10$^{20}$
cm$^{-2}$ with $b=50$ km s$^{-1}$. For each column density the lower and upper
dashed/dotted lines indicate the shifts for $b=25$ and $b=75$ km s$^{-1}$.
The different plots are for Ly$\alpha$ velocity widths of 1800, 1500 and 1000 
km s$^{-1}$ (FWHM). 
Figure 7 also shows the amount of flux that remains after
H\,{\small I} absorption as a function of absorber offset, for H\,{\small I} 
column densities 
and Ly$\alpha$ widths the same as above.
The solid lines are for $b=50$ km s$^{-1}$, while 
lower and upper dotted lines now correspond to $b=75$ and $b=25$ km s$^{-1}$.
Figure 8 indicates the maximum shift of the Ly$\alpha$ peak that can
be produced by H\,{\small I} absorption systems of different column densities and Doppler
parameters.

We find from these simulations that, depending on the velocity width of Ly$\alpha$
emission line, H\,{\small I} absorption systems can produce
velocity shifts in the observed peak of Ly$\alpha$ emission of 100--600 km s$^{-1}$
for H\,{\small I} column densities of 10$^{17}$--10$^{20}$ cm$^{-2}$. 
Such velocity shifts are comparable with the velocity shifts observed in our
sample of radio galaxies.
Thus, the velocity shifts between Ly$\alpha$ and the higher ionization lines
could be explained if the majority of
HZRGs have strong (N(H\,{\small I}) $\sim 10^{17}$--$10^{20}$ cm$^{-2}$)
associated absorption.

However, the presence of associated H\,{\small I} absorption systems cannot be
the cause for all observed velocity shifts between the various emission lines.
Of the objects with significant velocity shifts, there are at least five
with known strong H\,{\small I} absorption in the Ly$\alpha$ line 
\cite{oji96b}, 
but also four that do not have strong associated H\,{\small I} absorption. 
These last objects actually have some of the largest observed significant 
velocity shifts between the emission lines.
Furthermore, the origin of the observed systematic shift of 
C\,{\small IV} with respect to He\,{\small II} and C\,{\small III}]
is unclear. Possibly, different velocity profiles cause the observed shifts between these 
emission lines 
if they originate from different regions in the galaxies. We note that 
C\,{\small IV} 
is also sensitive to absorption. However, for an H\,{\small I} absorption with 
high column density ($>10^{18}$ cm$^{-2}$),
the associated C\,{\small IV} 
absorption column density is usually a factor 10$^4$--10$^5$ 
lower (e.g. Bergeron \& Stasi\'nska 1988)\nocite{ber86} and
will not have a very strong effect on the position of the peak of the line.

Thus, H\,{\small I} 
absorption systems cannot be the only cause of velocity shifts.
Additional mechanisms such as outflow or inflow combined with obscuration,
as proposed for quasars, or radio source induced displacements must be 
important.

\subsection{Continuum alignment of the $z>1.9$ Leiden compendium}
It is well known that the major axes of the optical continuum and the radio 
emission of radio galaxies
at $z>0.8$ are aligned, while low redshift radio galaxies
have a weak tendency to have the optical minor axis aligned with the radio
axis \cite{cha87,mcc87a}. Interpretations of the ``alignment effect''
include jet-induced star formation and scattering of anisotropically
emitted nuclear continuum \cite{beg89,ree89a,ser89b,tad92}.

The $R$ band continuum of 
many of the $z \geq 0.8$ radio galaxies is contaminated by strong emission
lines extended in the direction of the radio axis.
Such strong extended emission lines might cause
the alignment effect, which is often measured at the lowest
flux levels of the extended emission. It has therefore been suggested that
the observed alignment does not involve the optical continuum (e.g. 
Meisenheimer et al. 1994).\nocite{mei94}
At redshifts of $\sim 1$, the strong
[O\,{\small II}]$\lambda$3727 emission line is in $R$ band. This emission line is usually
spatially extended in the direction of the radio axis (see McCarthy et al. 
1991b).\nocite{mcc91a}
At redshifts between 1.9 and 3, however, there are no strong emission lines
in $R$ band. The only significant emission line present in $R$ band 
at these redshifts might be C\,{\small III}]$\lambda$1909. 
From the measured equivalent
widths measured for our sample of HZRGs we deduce that C\,{\small III}] 
contributes
less than 10\% to the $R$ band flux and often less than 5\%. 
Although C\,{\small III}] may be slightly spatially
resolved, it is very unlikely that it would cause a strong alignment.

For 18 of the $z>1.9$ objects of the Leiden USS compendium
we have the radio position angle from 20cm VLA
maps \cite{rot94} and an $R$ band image from which we could reliably
determine the optical position angle \cite{rot95}.
Figure 9 shows the difference between the optical continuum axis
and the radio axis for these 18 objects, divided into two groups with
less than 5\% line contamination in $R$ band and with line contamination 
between 
5\% and 10\%. The alignment is similarly strong for both subsets, indicating
that the minor contamination by line emission from C\,{\small III}] 
in $R$ band does not play a role. This plot provides further
evidence that the previously inferred alignment between the optical continuum
of HZRGs and the radio structure is real.

\section{Radio source sizes}

The observed differences between steep spectrum quasars and radio
galaxies have been explained as merely due to whether the radio source
axis is more closely aligned with the line of sight (quasars) or in
the plane of the sky (galaxies)(see Antonucci 1993 \nocite{ant93} and
references therein).  Clearly this model predicts that the projected
linear sizes of quasars should on average be smaller than that of
radio galaxies.  Barthel (1989) \nocite{bar89} found that steep-spectrum
quasars with $0.5 < z< 1$ have sizes that are systematically smaller
than radio galaxies with $0.5 < z< 1$ by a factor 2.2, indicating 
a cone angle of $\sim 45\deg$ along which the quasar can be seen.
Gopal-Krishna and Kulkarni (1992) \nocite{gop92} have obtained a similar
results for a sample of radio sources with $0.1 < z < 2 $. At low
redshift, it seems that there is a relative shortage of observed steep
spectrum quasars \cite{kap90c}, indicating that the simple unification
scheme with a fixed cone angle breaks down at low $z$.

We have compared the radio source size distribution for the $z>2$ USS
radio galaxies presented here with the radio source sizes for the
$z>2$ quasars from the quasar sample studied by Barthel and Miley
\cite*{bar88b} (see Fig. 10). In both cases the linear sizes have been
calculated using a Hubble constant of $H_0 = 75 $ km s$^{-1}$
Mpc$^{-1}$ and a decelaration parameter of $q_o=0.5$.  The mean
angular radio size of the quasars is $ 18 \pm 6 $ and of the USS radio
galaxies $40 \pm 7.5$.  The ratio between the sizes is 2.2, the same
factor as has been found for the lower redshift objects.  Within the
context of the model of Barthel, this would imply that the cone angle
of $\sim 45\deg$ that had been found for the sources with $0.5 < z <1$
also holds at high redshift.

However, there are two selection effects.  First, as we have pointed
out previously, there is an angular size bias in the USS sample in
favour of small radio angular sizes.  Secondly, the radio galaxies all
have have been selected to have ultra-steep spectral indices, whereas
the quasars form a proper flux limited sample with a median
spectral index of $-0.8$.  We show the projected linear size of the
radio galaxies and quasars as a function of spectral index in Fig. 11.
This shows evidence that the steepest-spectrum quasars tend indeed to
be the largest. Restricting the sample to USS quasars, the 4 remaining
sources are too few for a meaning full statistical comparison with the
USS radio galaxies.

We therefore conclude that the difference in the source size
distribution for the $z>2$ USS radio galaxies and radio galaxies is
merely tentative. A proper comparison between the sizes of radio
galaxies and quasars should include data covering a similar range of
spectral indices for each species.

\section{Summary and conclusions}
In this paper we have presented spectroscopic results 
for galaxies associated with 
64 ultra steep spectrum radio sources. Of these radio galaxies, 
29 are at $z>2$ and the three highest redshifts are 3.6, 3.8 and 3.8.
Our ultra steep spectrum (USS) criterion ($\alpha < -1$) has proven to be the
most efficient way of finding distant radio galaxies.
We find that even among the USS sources,
there is a strong correlation between the spectral index 
and redshift. The most distant radio galaxies within the USS sample 
have the steepest radio spectra.

In our sample there are 3 radio galaxies at $z>3$ and 26 at $2 < z < 3$.
However, the present sample of USS radio galaxies does not allow us to decide 
whether there is a decrease in co-moving source density at the highest 
redshifts, because of the uncertain flux limit of the sample of radio sources.

We have analyzed the spectra of the 30 objects with the highest redshifts
($z>1.9$). For these high redshift radio galaxies, Ly$\alpha$ is almost always the
dominant emission line, with a rest frame equivalent width ranging from
$\sim100$ \AA\ to more than 1000 \AA. The equivalent widths of the most
important emission lines (Ly$\alpha$, C\,{\small IV}, He\,{\small II}, 
C\,{\small III}]) are found to correlate
strongly with each other. The large rest frame equivalent widths and the
correlation between the equivalent widths of the emission lines, confirm that
photoionization by a central continuum source is most likely the dominant
ionization mechanism.

There are significant velocity differences between the 
various emission
lines of our high redshift radio galaxies; 
mainly the Ly$\alpha$ line is shifted with
respect to the higher ionization lines. Velocity shifts range from 100 to
almost 1000 km s$^{-1}$ in some cases.
Simulations show that the effects of associated H\,{\small I} absorption on the
Ly$\alpha$ emission line may be responsible
for most of these velocity shifts. However, in at least
a few objects, other mechanisms such as organized kinematics of
the Ly$\alpha$ emission line gas (e.g. inflow or outflow) and obscuration
of the line emission from the far side of the galaxy must play a role.

The sample of objects discussed here is well suited for follow-up observations.
Various detailed studies of their properties are presented in other papers
or are underway, both with ground-based telescopes and with the HST.

\section*{Acknowledgements}

We acknowledge support from an EU twinning project, funding from the
high-z programme subsidy granted by the Netherlands Organization for
Scientific Research (NWO) and a NATO research grant. The work by WvB
for this project was performed at IGPP/LLNL under the auspices of the
U.S. Dept. of Energy under contract W-7405-ENG-48.

\section{Figures} 

\begin{figure} 
\vspace{-1cm} 
\centerline{ 
\psfig{figure=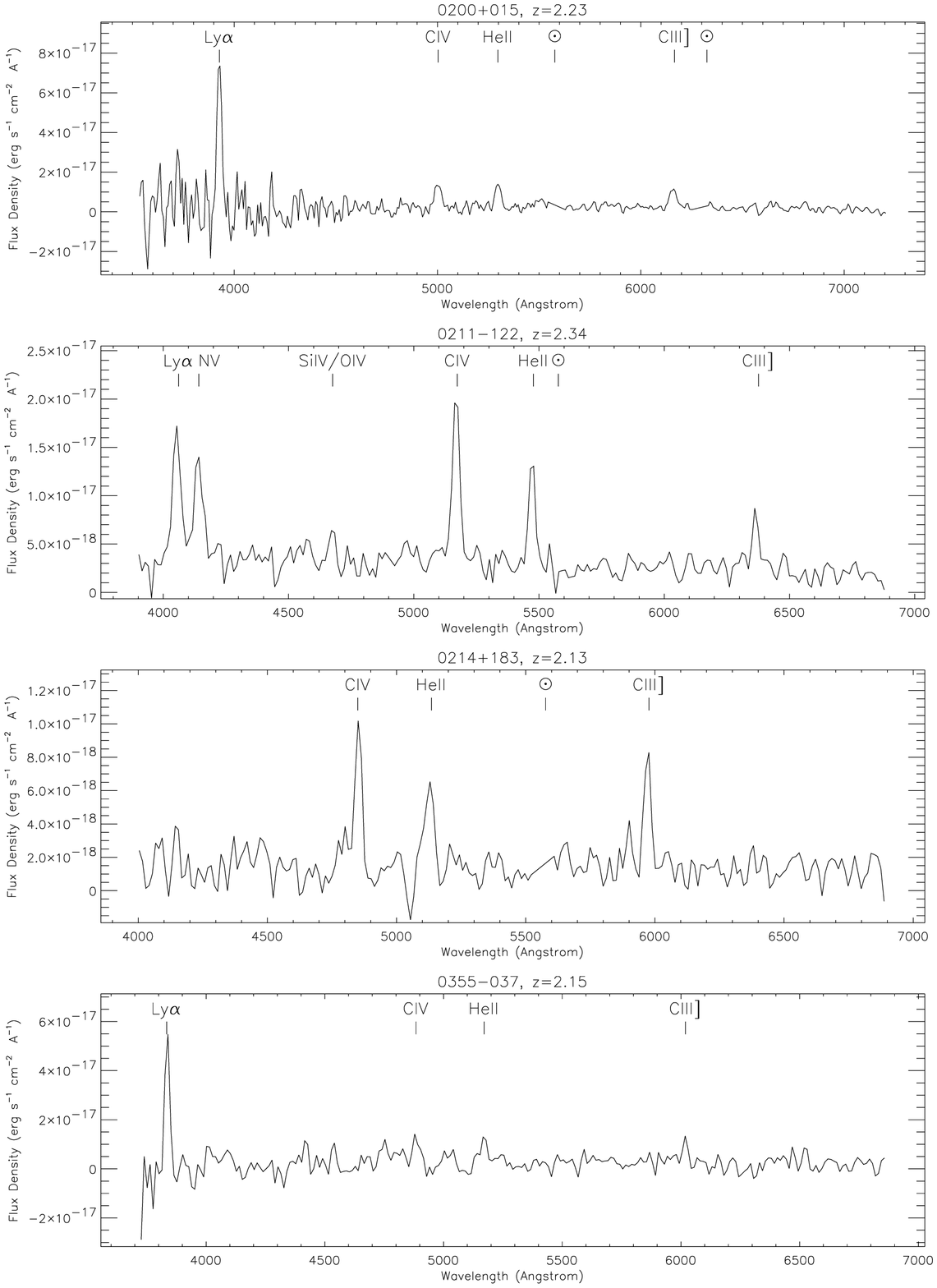,height=21cm} 
}
\noindent{\bf Fig. 1.} One dimensional spectra of $z>1.9$ USS sources 
\end{figure} 

\newpage \clearpage

\begin{figure}
\vspace{-1cm} 
\centerline{ 
\psfig{figure=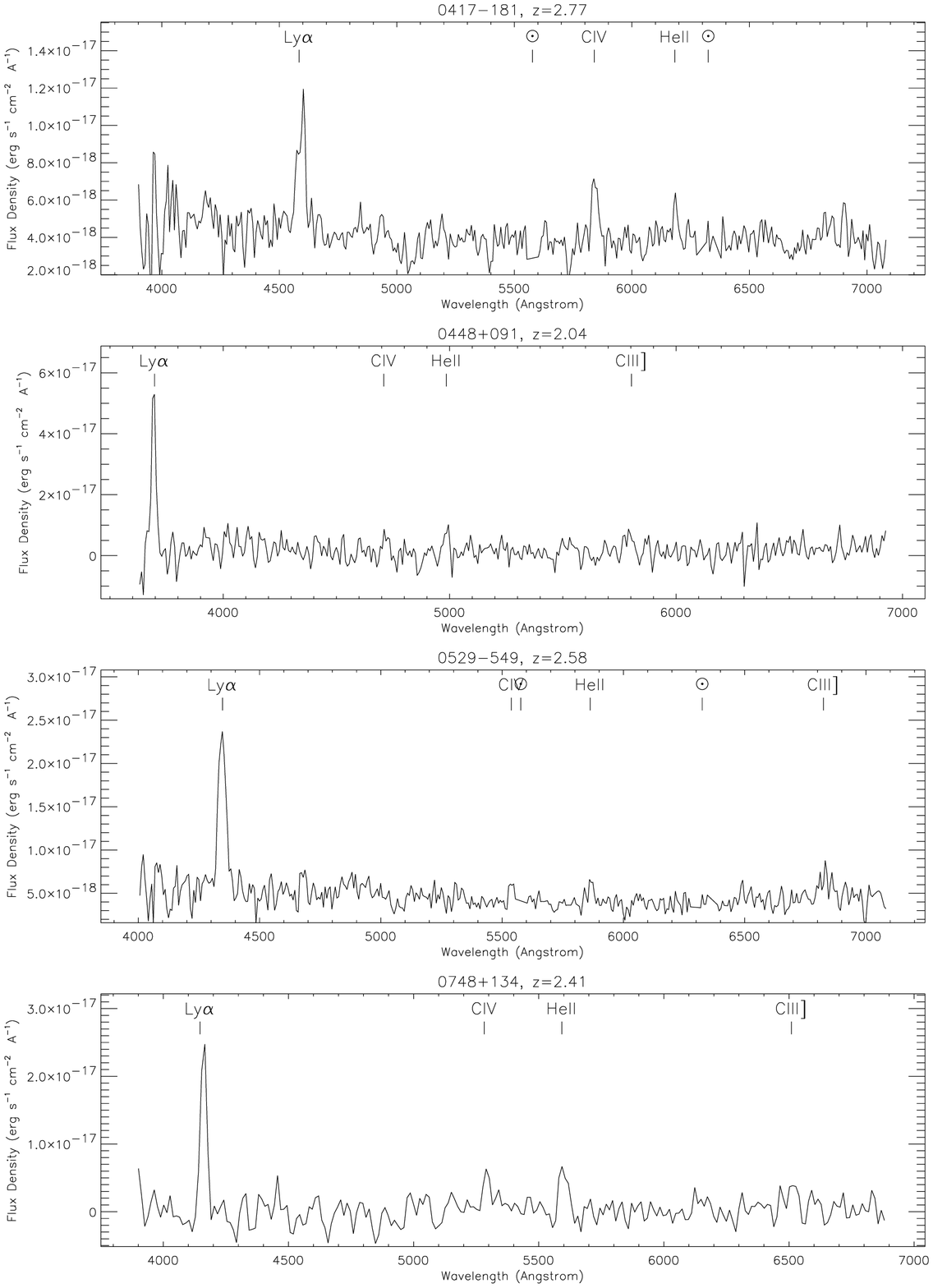,height=21cm} 
}
\noindent{\bf Fig. 1.} -- continued --. 
\end{figure} 

\newpage \clearpage  

\begin{figure}
\vspace{-1cm}
\centerline{ 
\psfig{figure=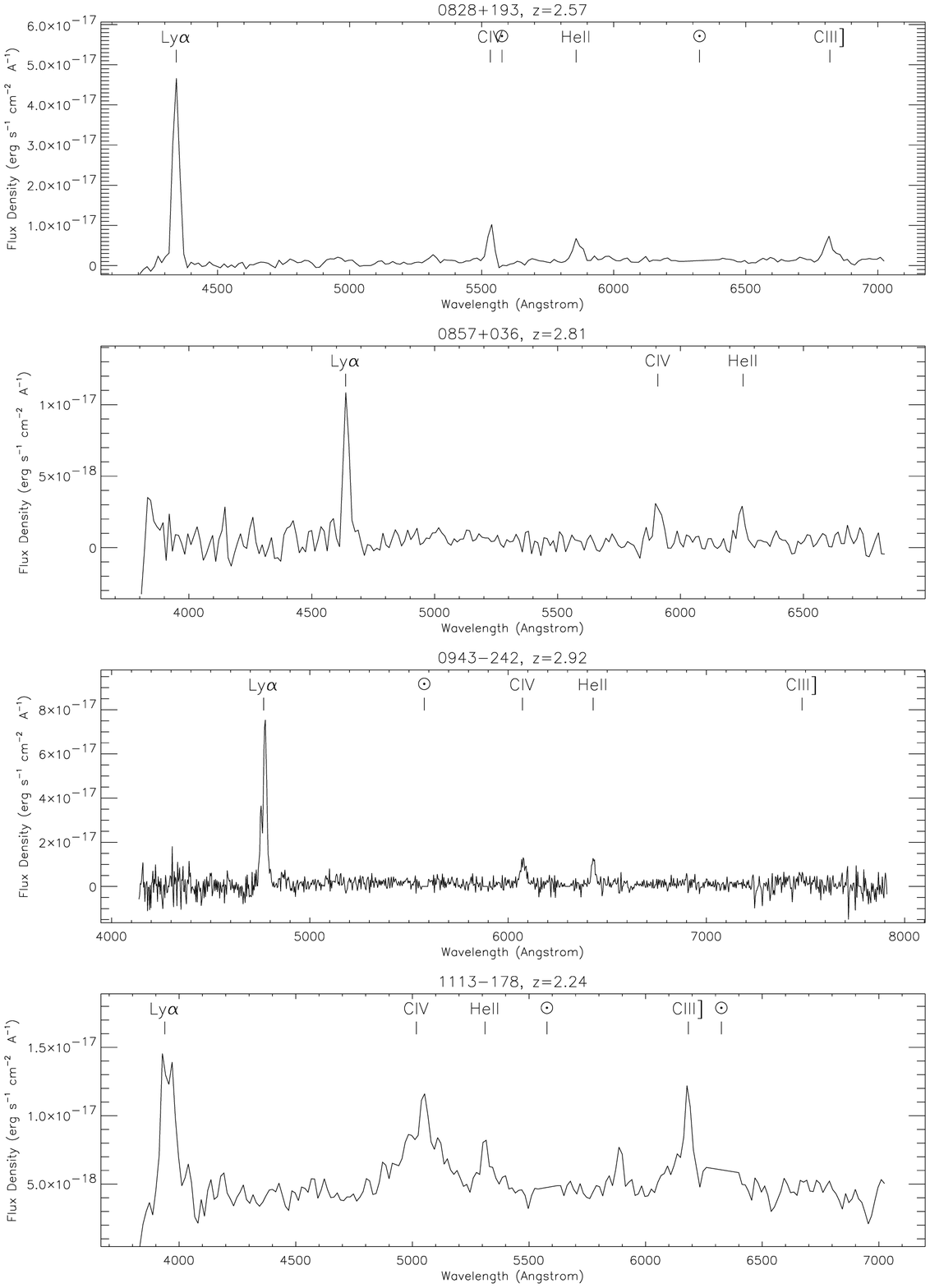,height=21cm} 
}
\noindent{\bf Fig. 1.} -- continued --. 
\end{figure} 

\newpage \clearpage  

\begin{figure}[hp]  
\vspace{-1cm}
\centerline{ 
\psfig{figure=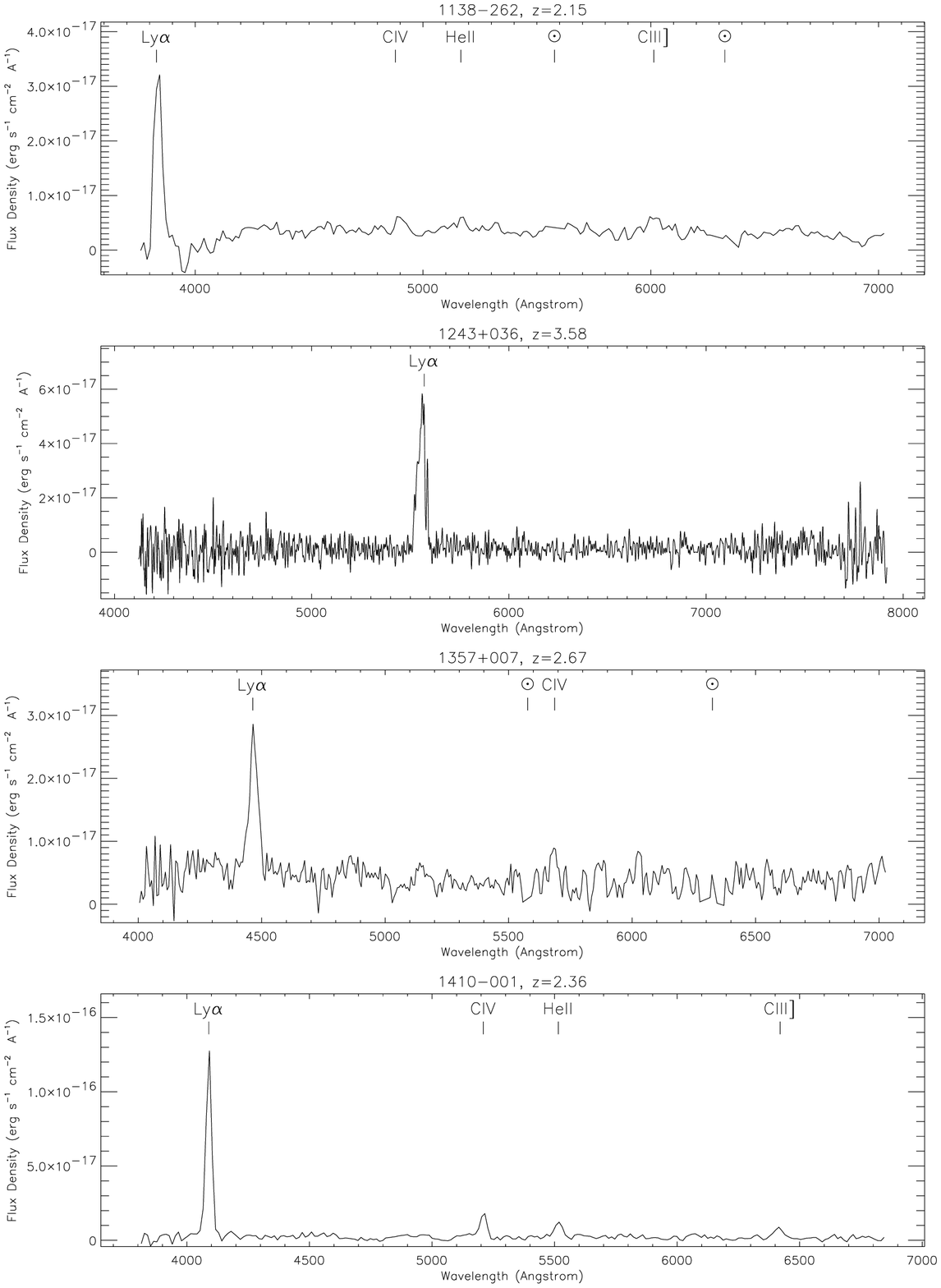,height=21cm} 
}
\noindent{\bf Fig. 1.} -- continued --. 
\end{figure} 

\newpage \clearpage 

\begin{figure}[hp]  
\vspace{-1cm}
\centerline{ 
\psfig{figure=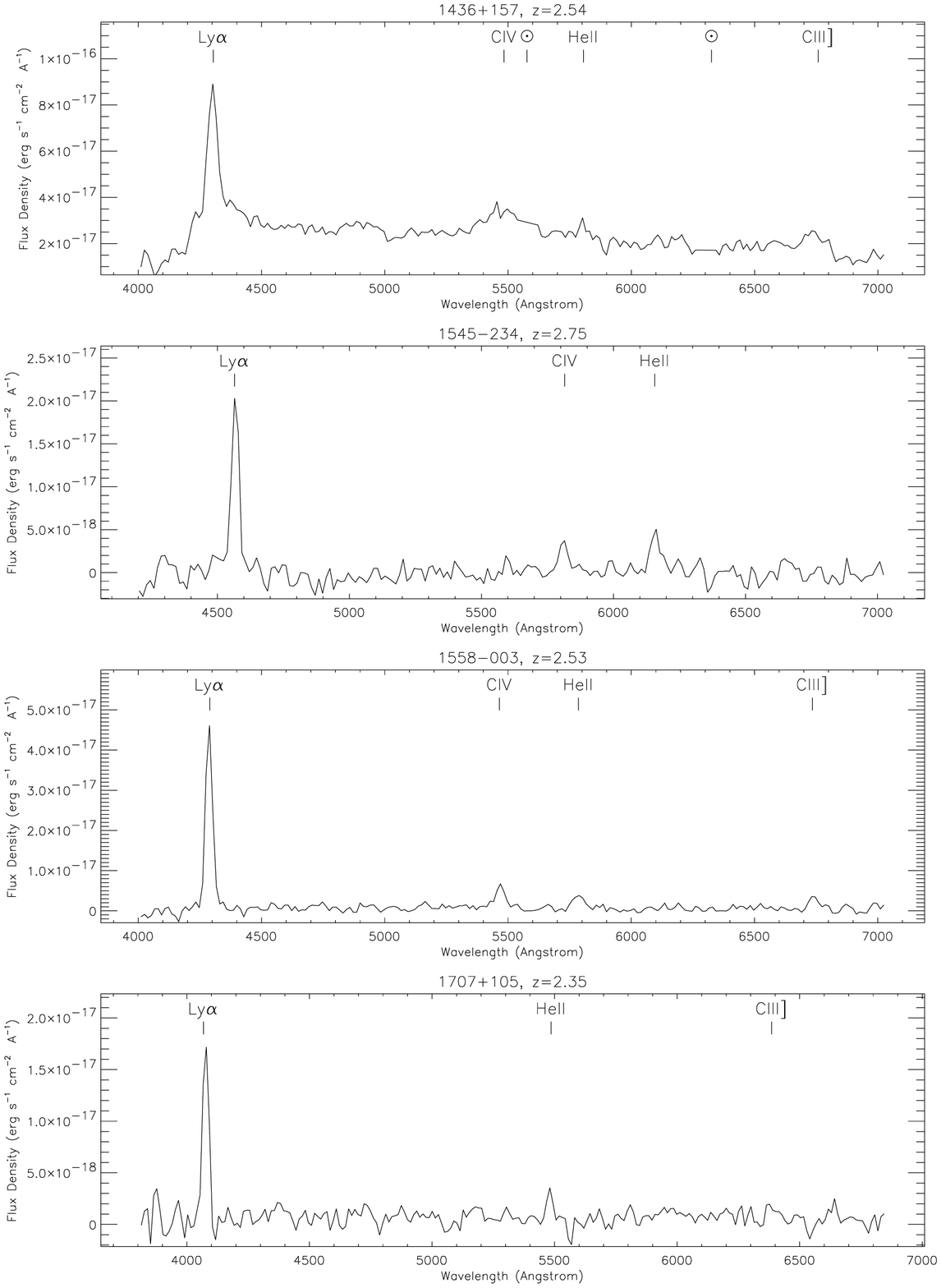,height=21cm} 
}
\noindent{\bf Fig. 1.} -- continued --. 
\end{figure} 
\newpage \clearpage 

\begin{figure}[hp]  
\vspace{-1cm}
\centerline{ 
\psfig{figure=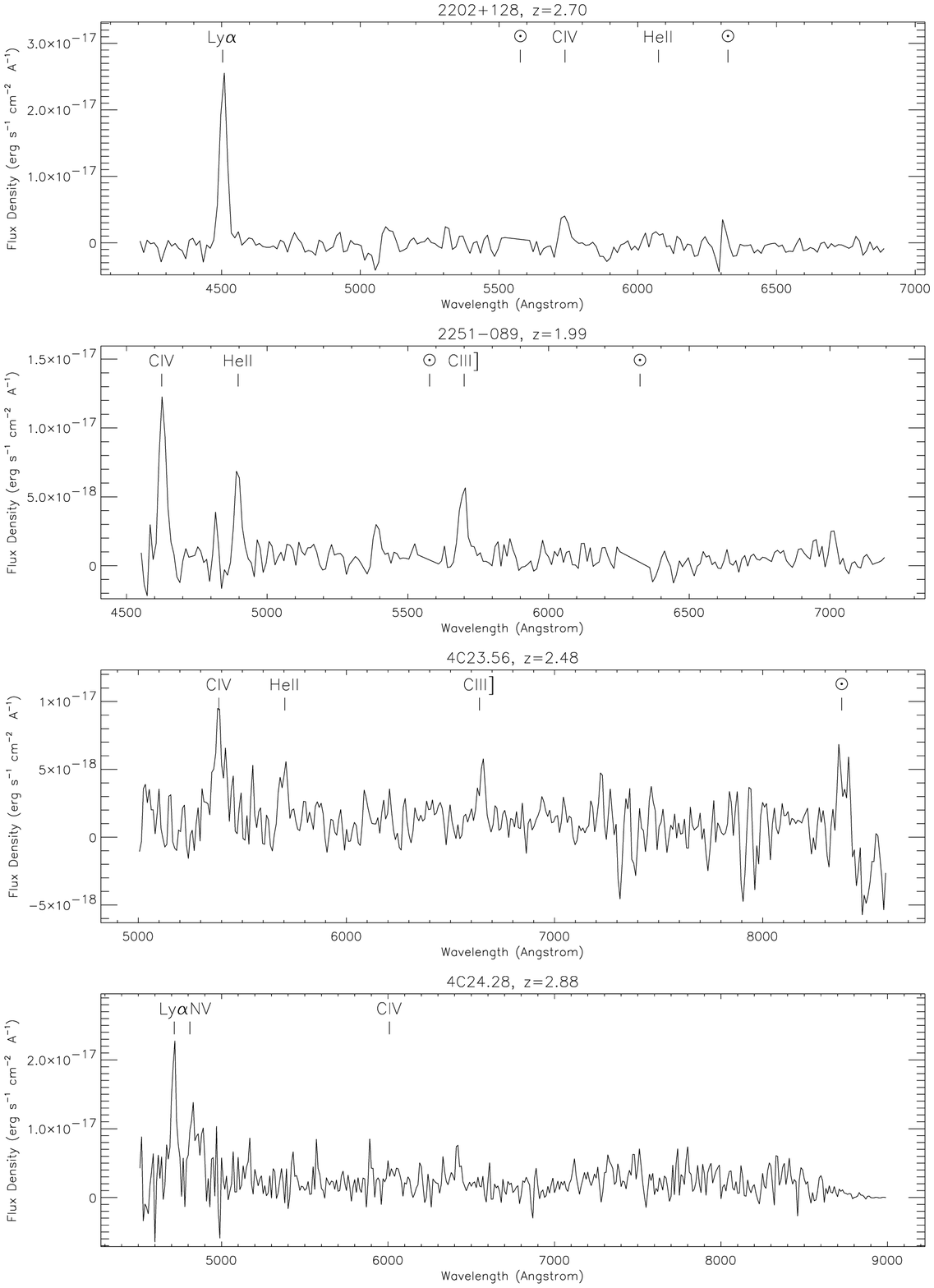,height=21cm} 
}
\noindent{\bf Fig. 1.} -- continued --. 
\end{figure} 

\newpage \clearpage

\begin{figure}[hp]  
\vspace{-1cm}
\centerline{ 
\psfig{figure=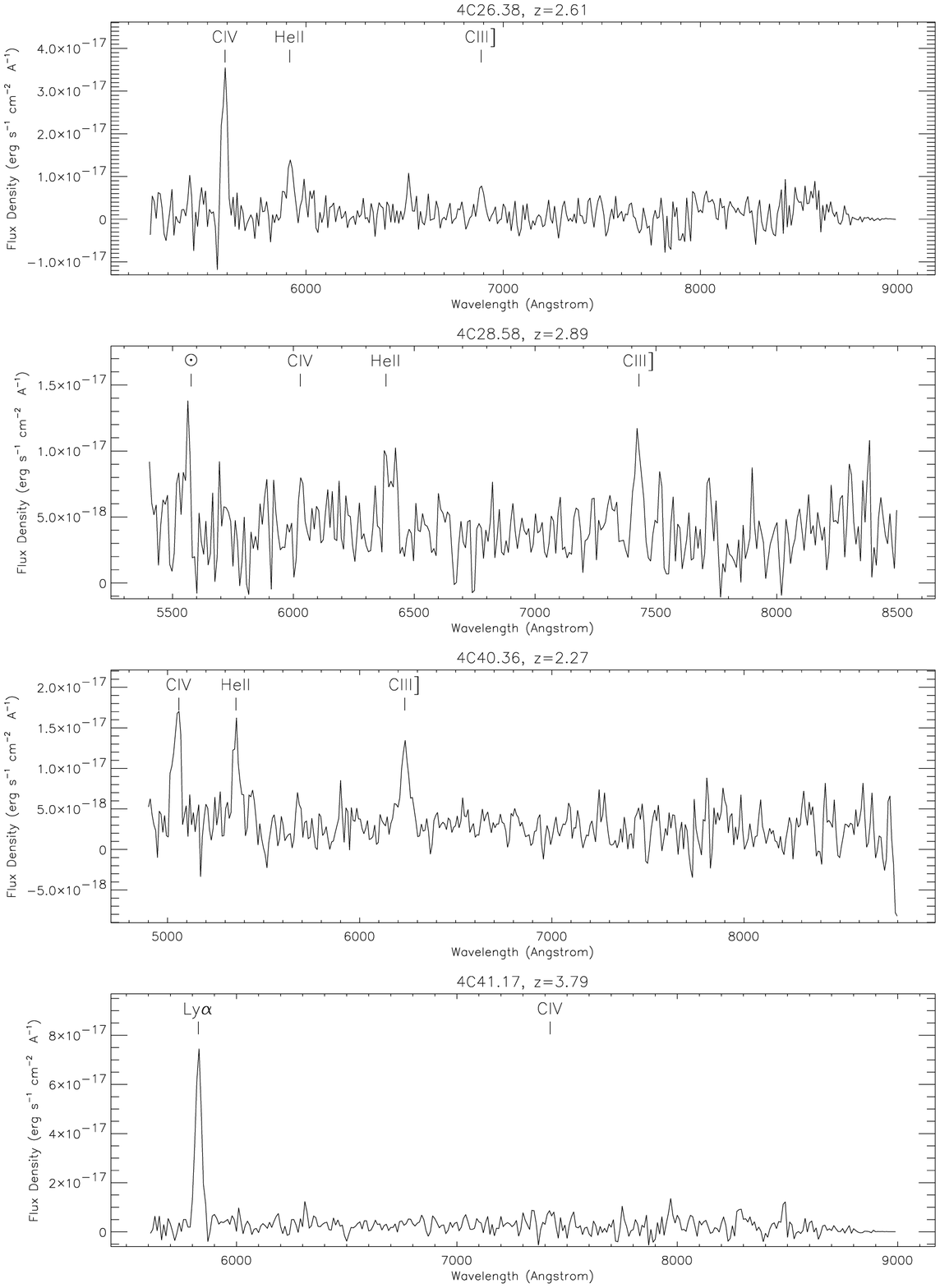,height=21cm} 
}
\noindent{\bf Fig. 1.} -- continued --. 
\end{figure} 
\newpage \clearpage

\begin{figure}[hp]  
\vspace{-1cm}
\centerline{ 
\psfig{figure=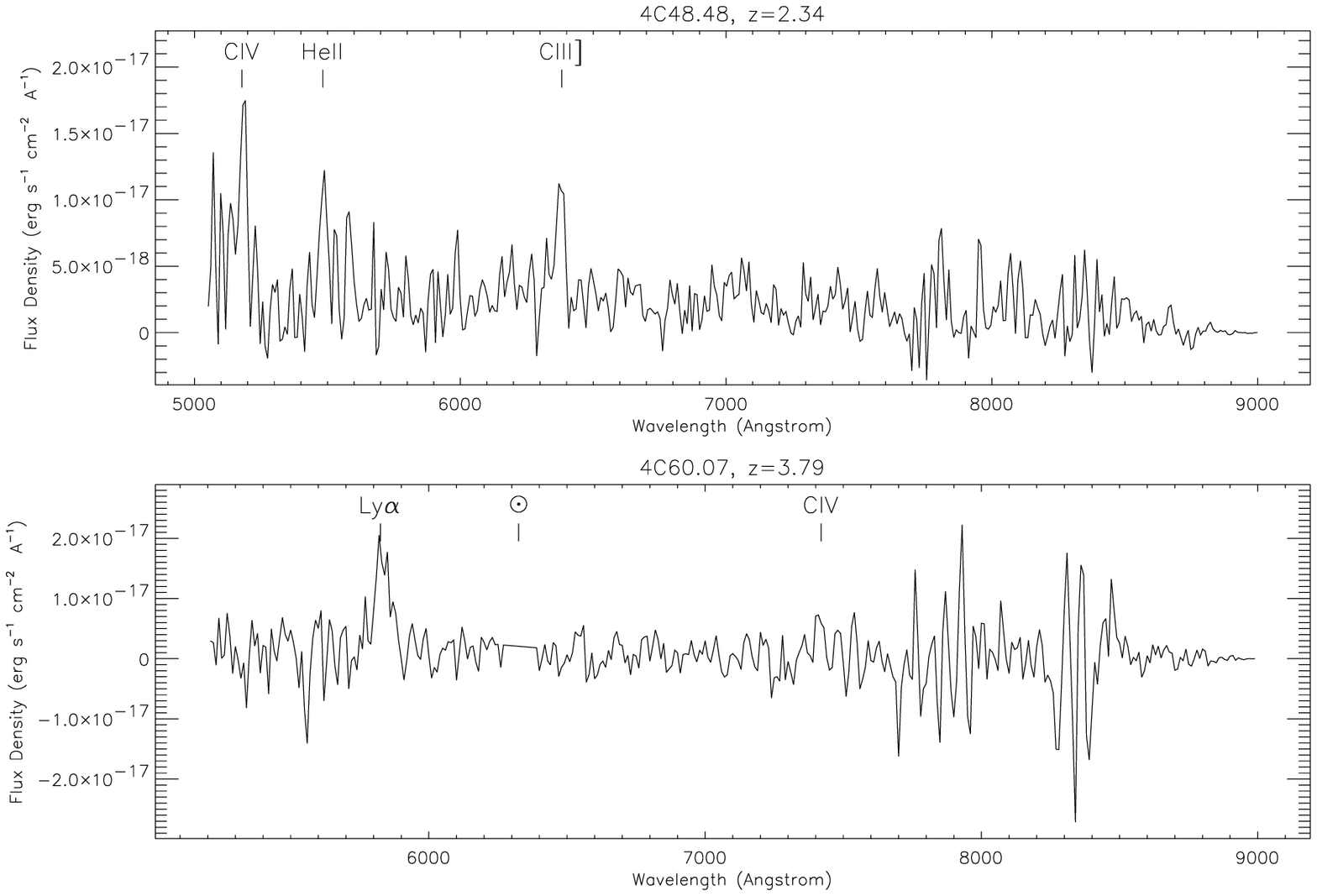,height=21cm} 
}
\noindent{\bf Fig. 1.} -- continued --. 
\end{figure} 

\newpage \clearpage 

\begin{figure}[p]
\hspace{1cm}
\hbox{
\psfig{figure=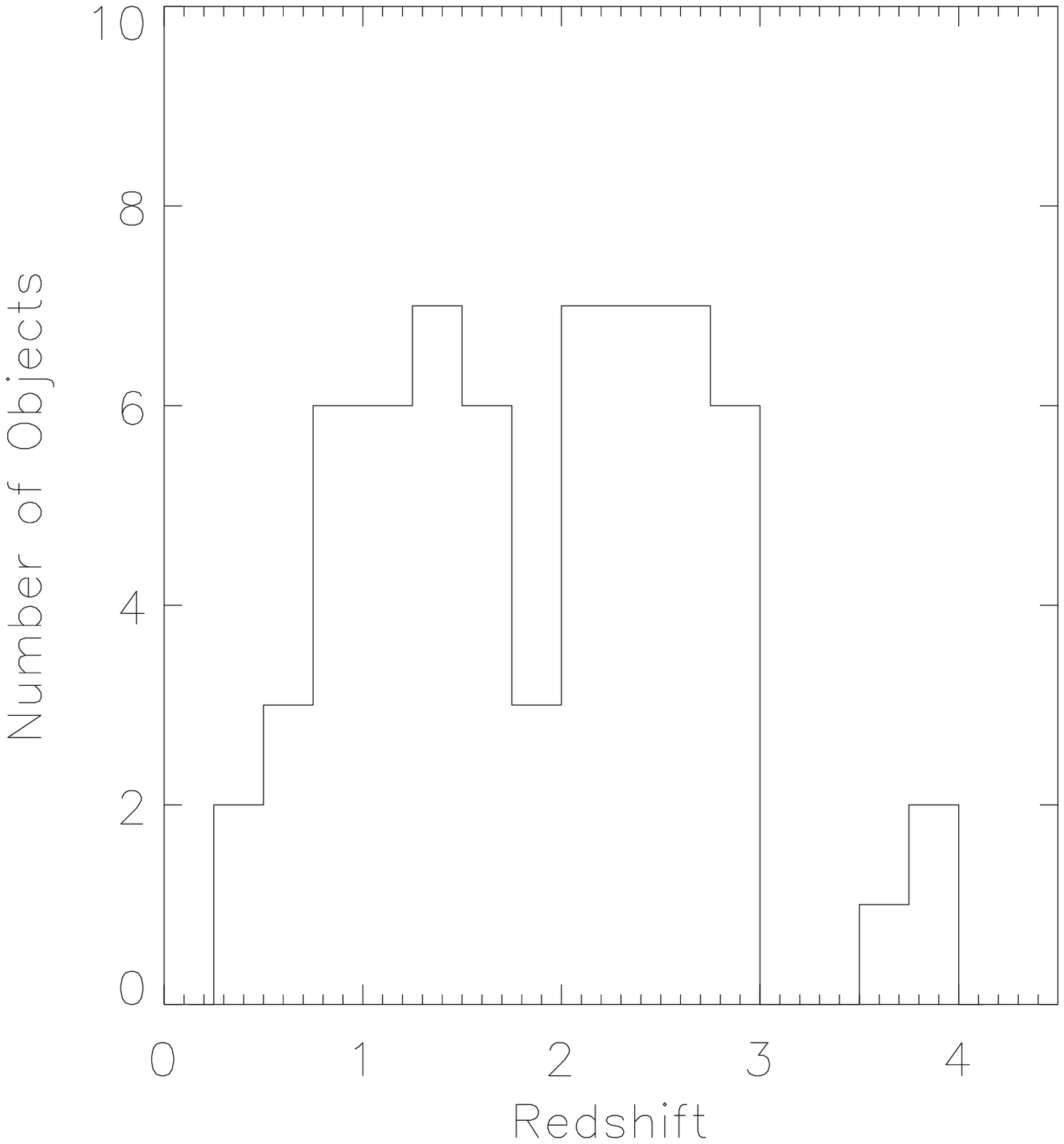,width=12cm,height=10cm}}

\noindent{\bf Fig. 2.} Redshift distribution of the USS radio galaxies
in our sample whose redshifts could be determined
\end{figure}

\newpage \clearpage

\begin{figure}[p]
\hbox{
\psfig{figure=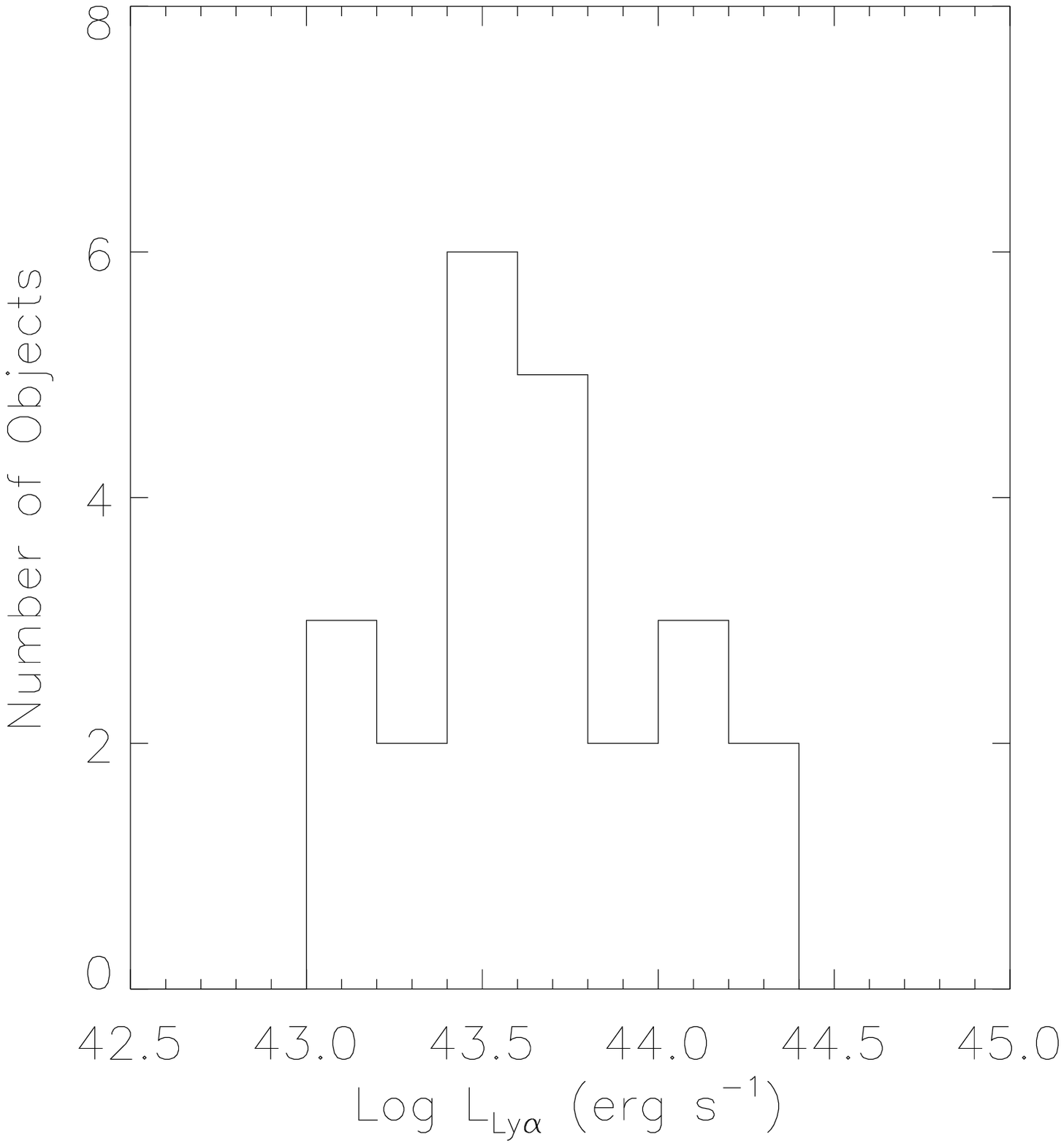,width=12cm,height=10cm}}
\noindent{\bf Fig. 3.} Distribution of the Ly$\alpha$ luminosities
\end{figure}
\newpage \clearpage

\begin{figure}[p]
\hbox{
\psfig{figure=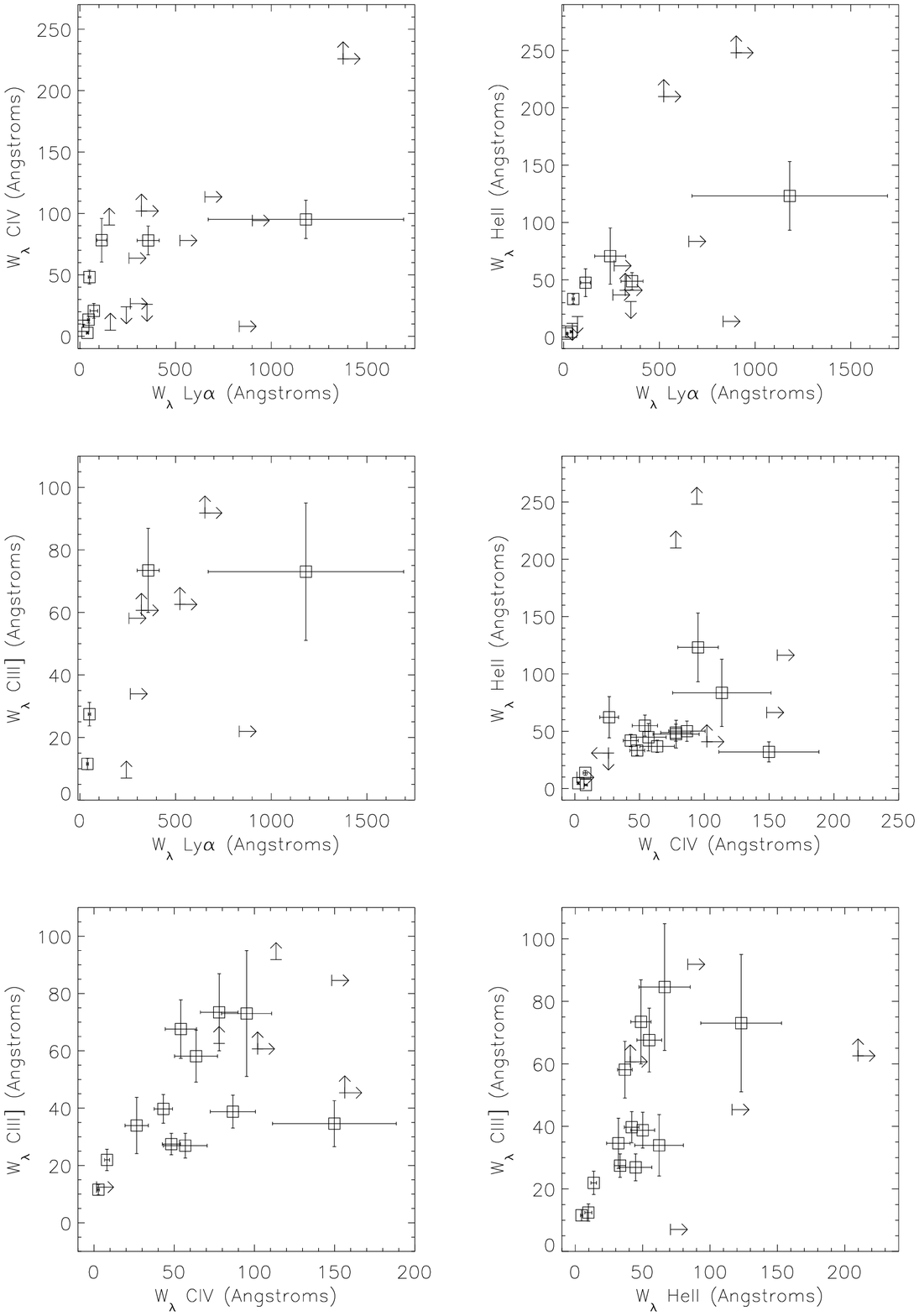,width=15cm}}
\noindent{\bf Fig. 4.} The rest frame equivalent widths of the
emission lines Ly$\alpha$, C\,{\small IV}, He\,{\small II} and 
C\,{\small III}] plotted against each other.
Lower limits are indicated in the plot as upward and rightward pointing
arrows, upper limits as downward and leftward pointing arrows
\end{figure}

\newpage \clearpage 

\begin{figure}[p]
\hbox{
\psfig{figure=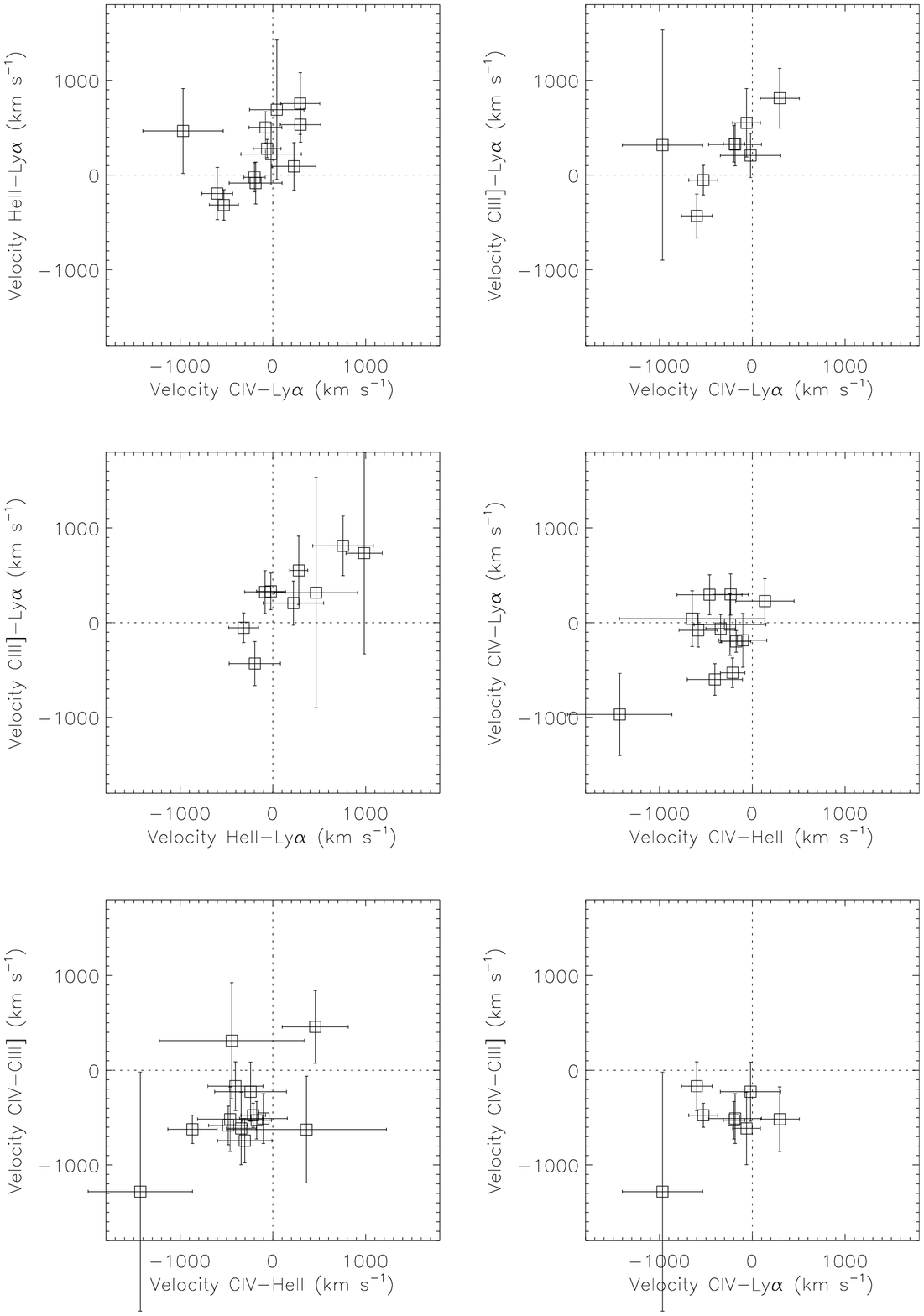,width=15cm}}
\noindent{\bf Fig. 5.} The velocity shifts between the 
emission lines Ly$\alpha$, C\,{\small IV}, He\,{\small II} and 
C\,{\small III}] plotted against each other
\end{figure}

\newpage \clearpage

\begin{figure}[p]
\centerline{
\psfig{figure=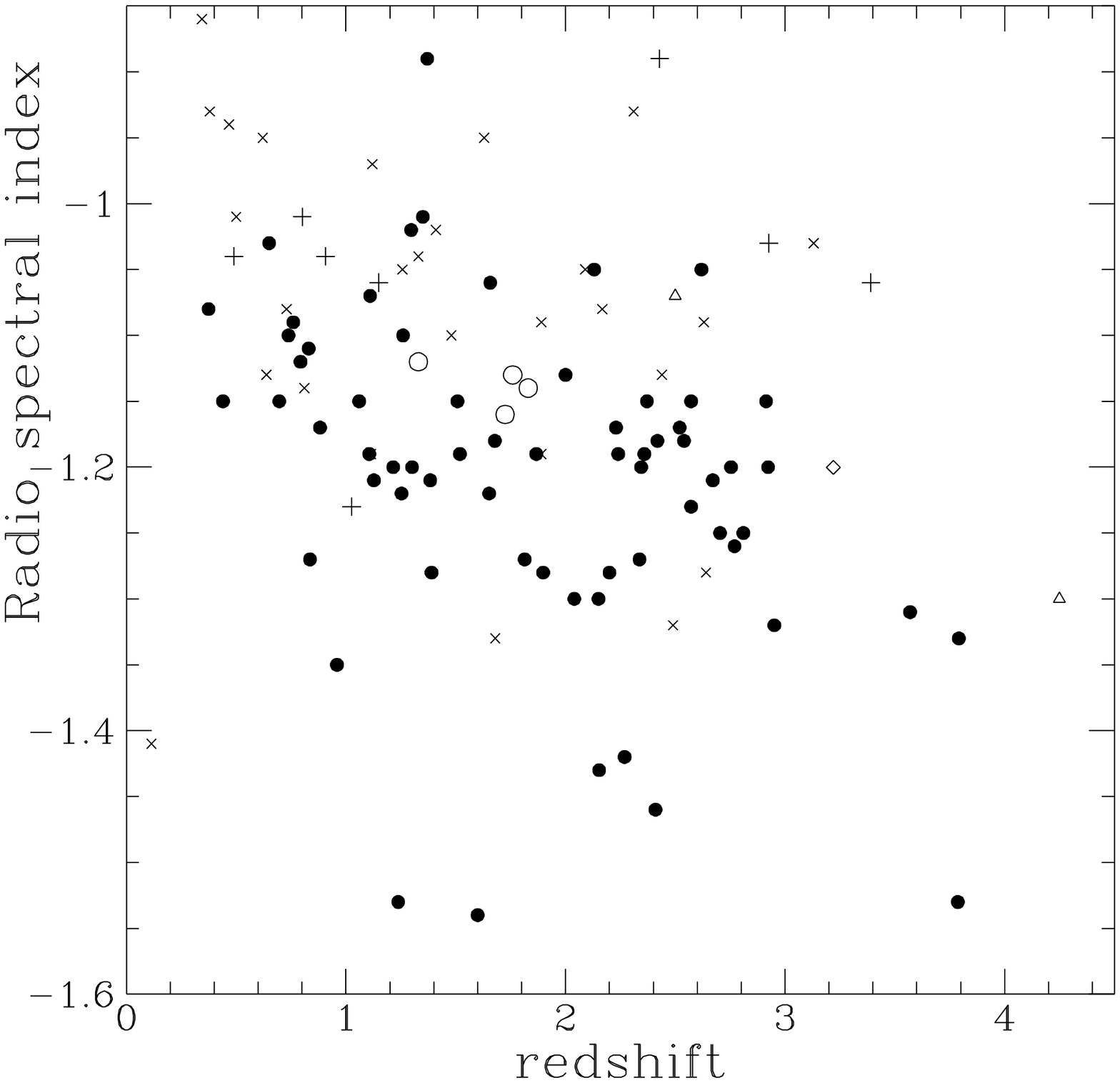,width=12cm}}

\noindent{\bf Fig. 6.} Radio spectral index plotted against redshift
for 108 radio galaxies. Filled dots are from our USS sample. The spectral
indices are determined between the lowest available frequency above 150 MHz and
5 GHz. The
open circles are also from our USS sample with redshifts obtained at 
Lick Observatory by McCarthy \& van Breugel (1994, private communication).
The crosses ($\times$) are from the Molonglo sample of McCarthy (1990a,1990b),
the pluses ($+$) are Bologna sources (McCarthy 1991, Lilly 
1988), the triangles are from the 8C survey
(Lacy et al. 1994, Lacy 1992) and the diamond is 6C\,1232+39 (Eales \& 
Rawlings 1993). 
\end{figure}

\newpage \clearpage 

\begin{figure}[p]
\vspace{-1.5cm}
\centerline{
\psfig{figure=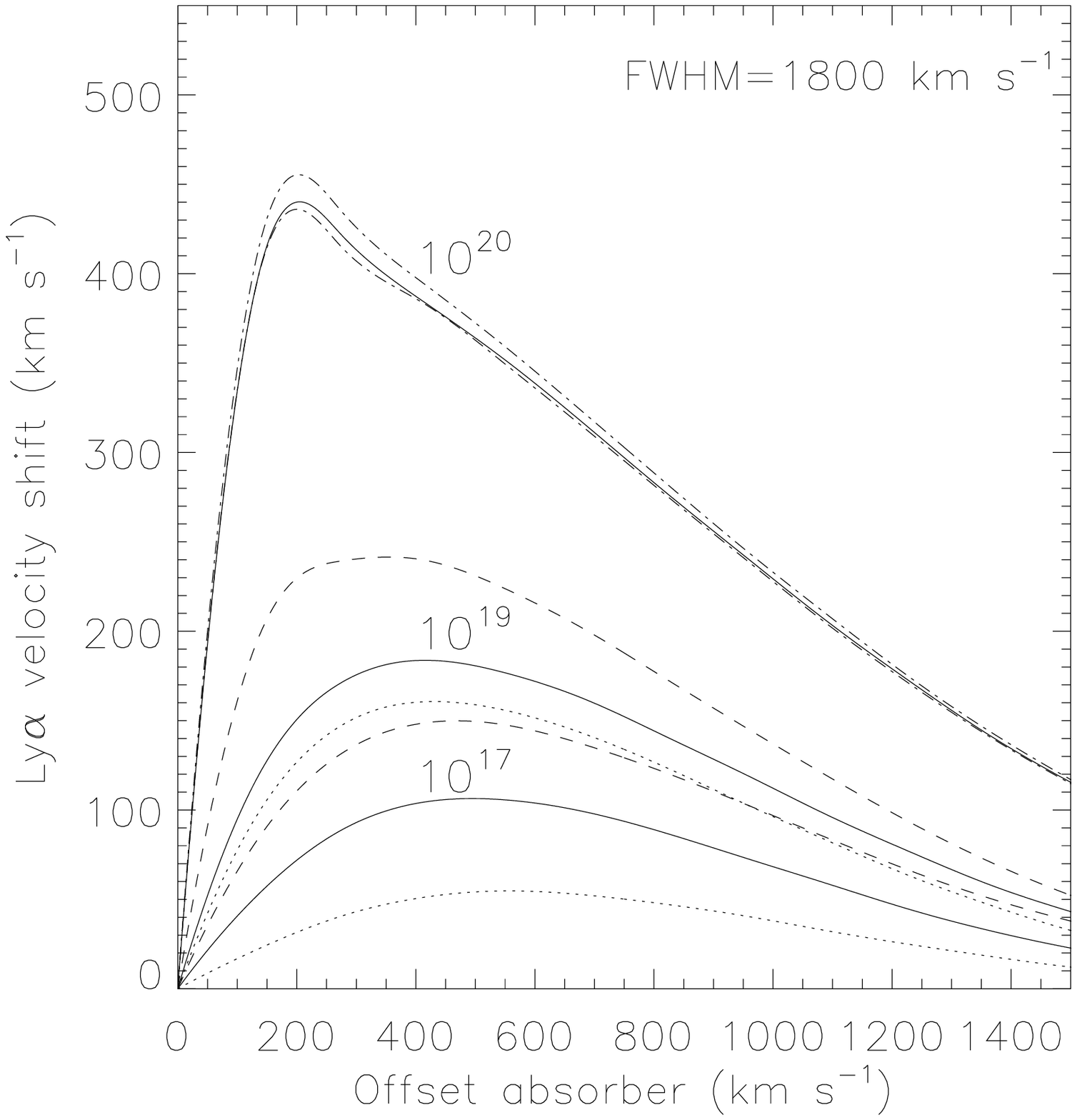,height=7.4cm}
\psfig{figure=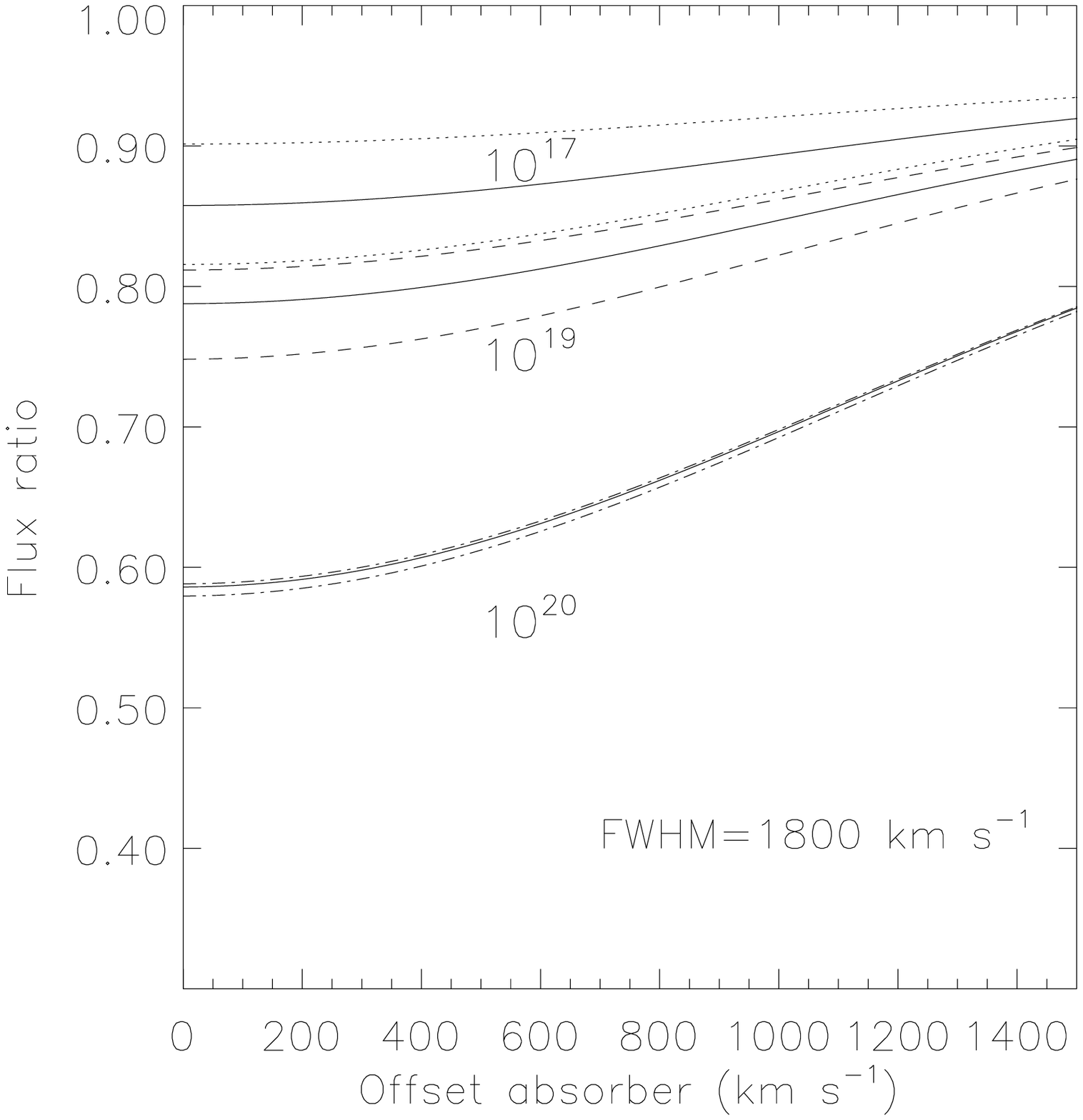,height=7.4cm}
}
\vspace{-0.6cm}
\centerline{
\psfig{figure=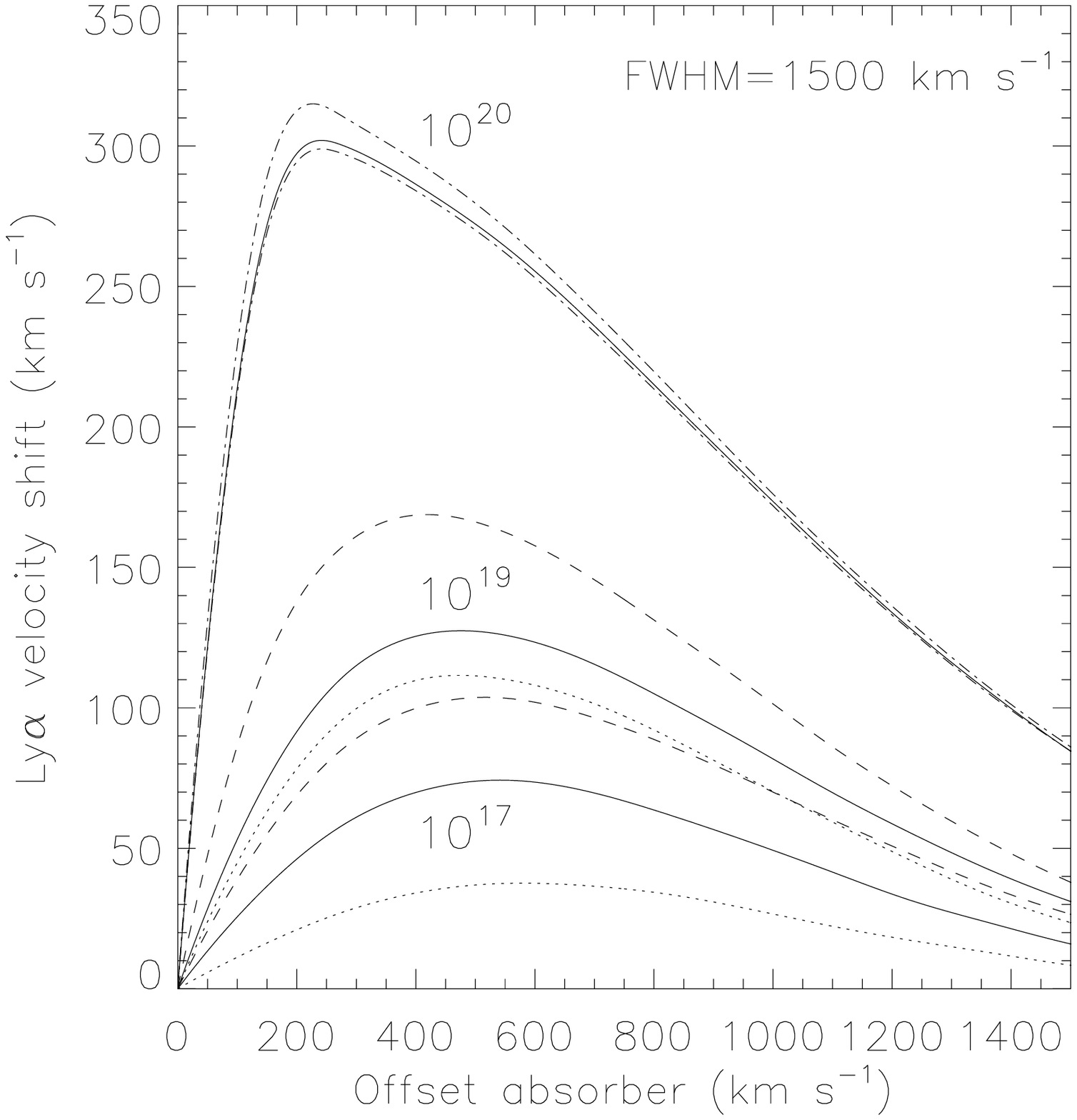,height=7.4cm}
\psfig{figure=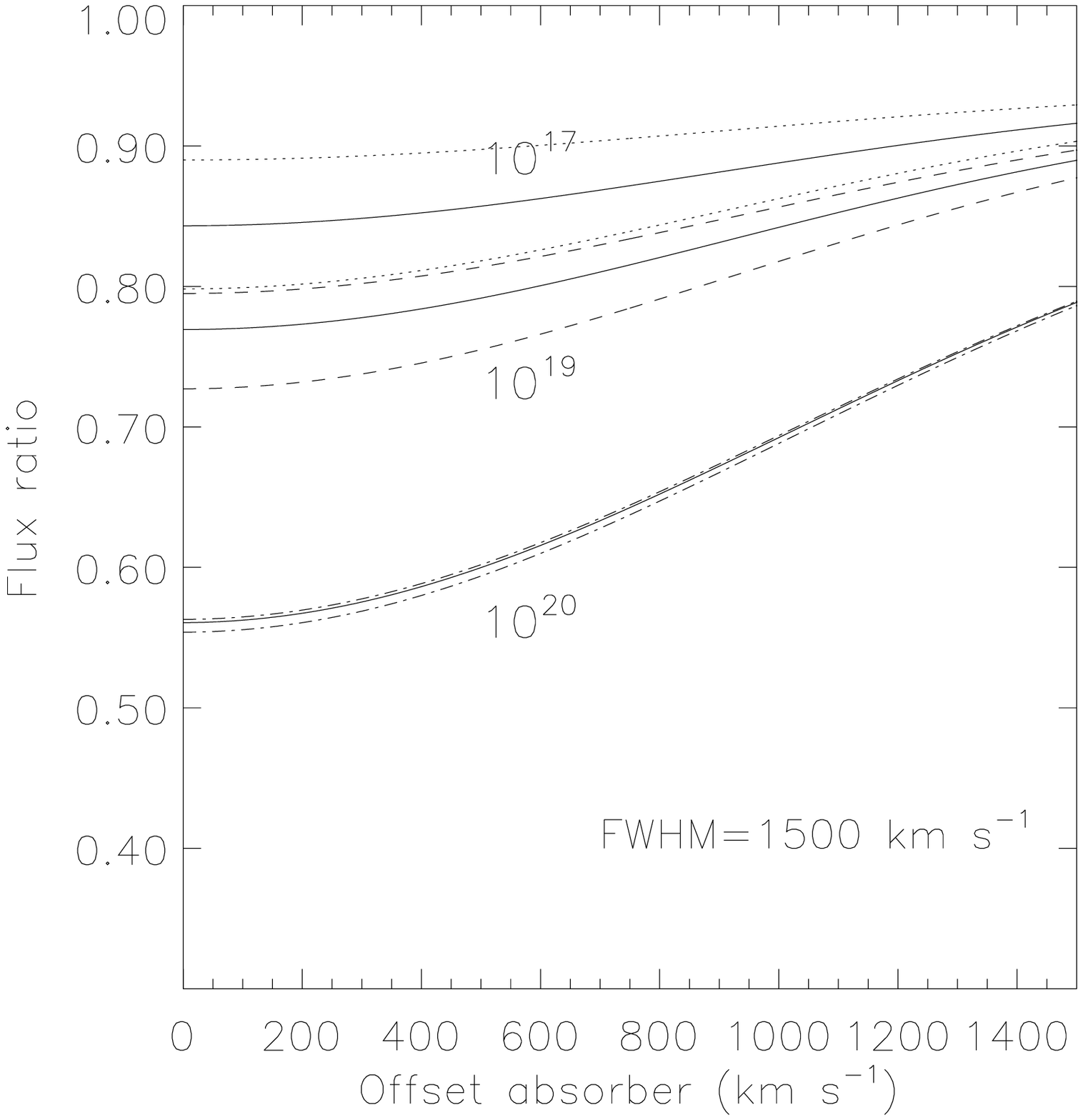,height=7.4cm}
}
\vspace{-0.6cm}
\centerline{
\psfig{figure=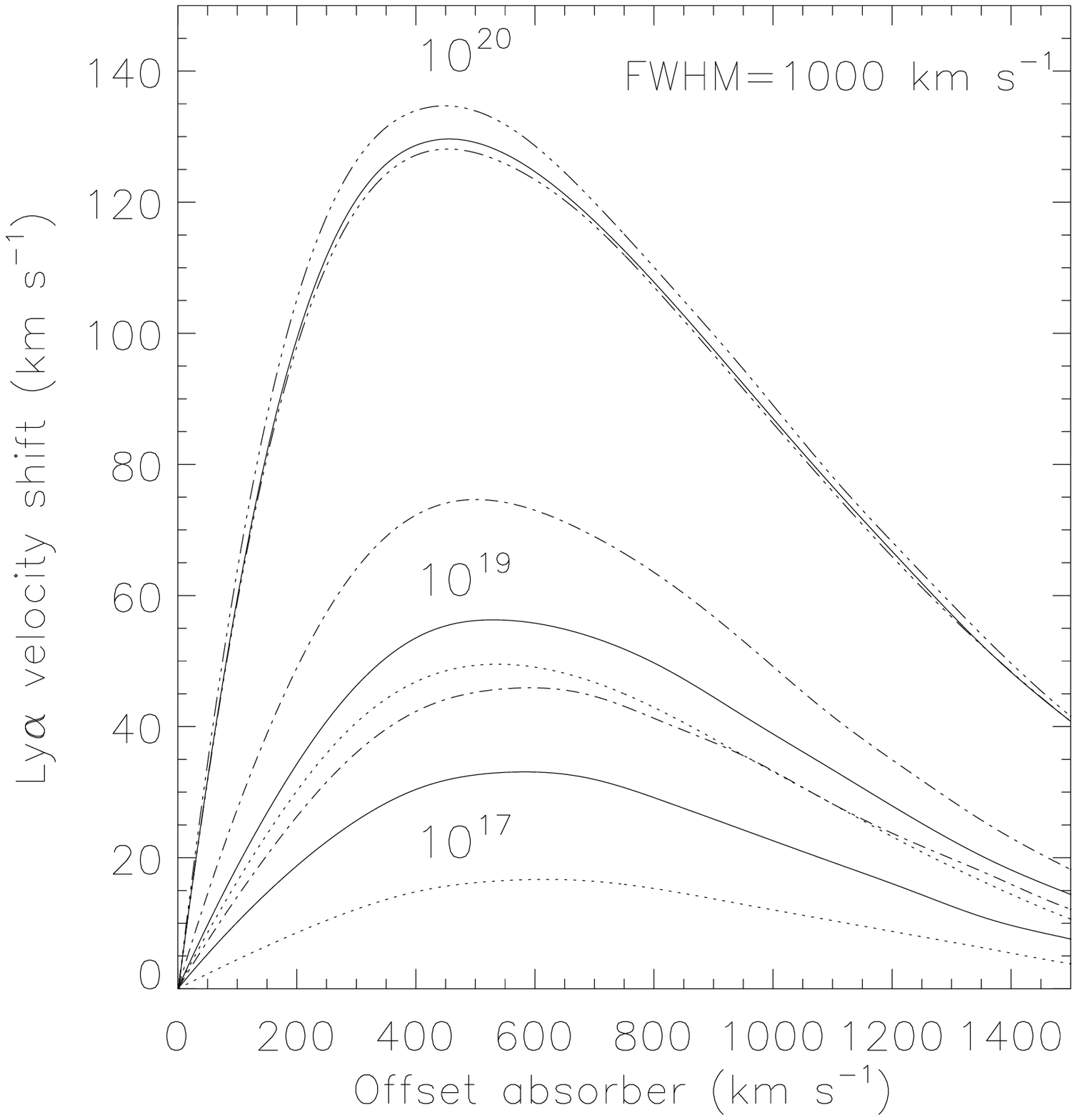,height=7.4cm}
\psfig{figure=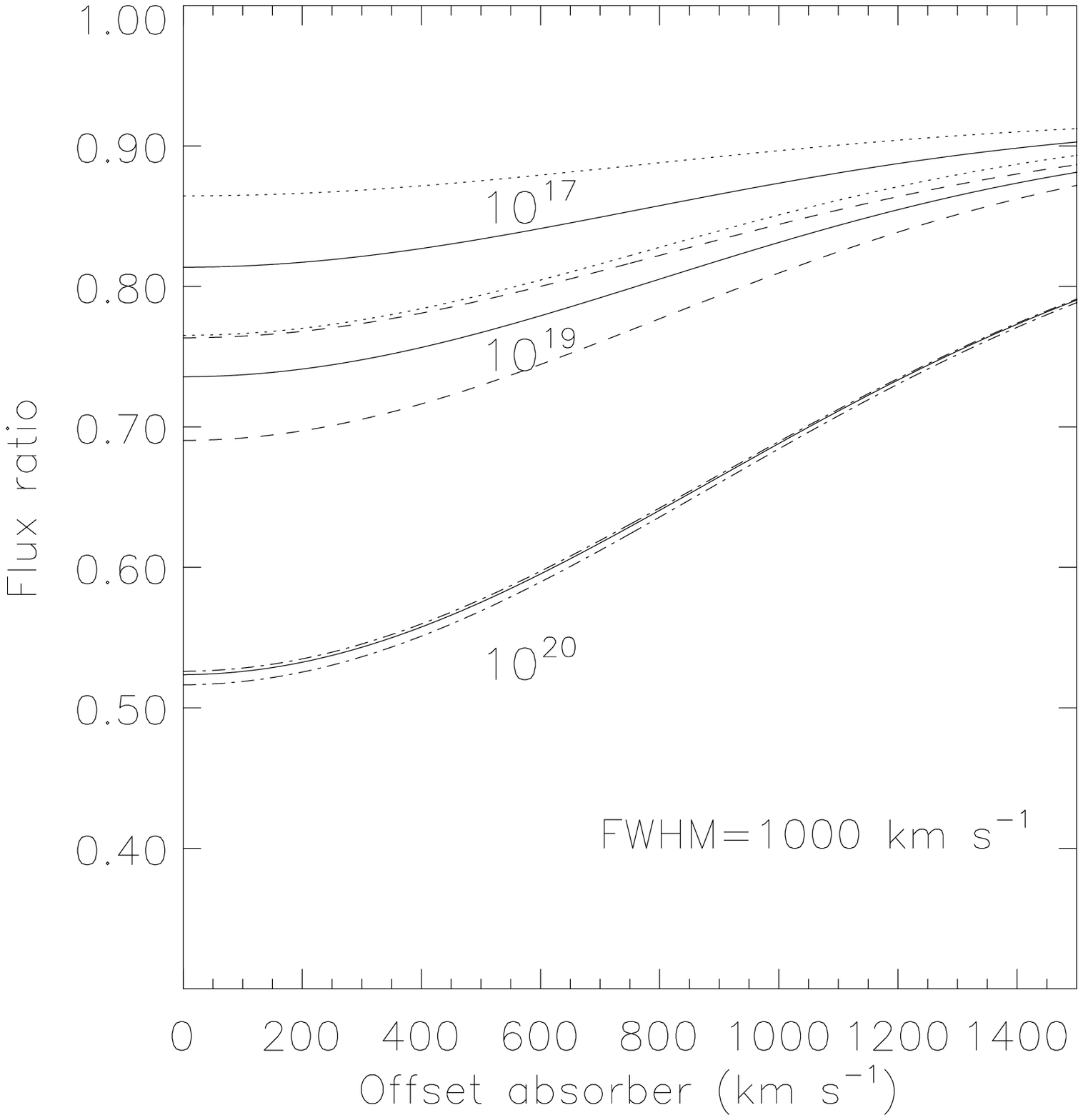,height=7.4cm}
}

\noindent{\bf Fig. 7.} {\it Left)} Velocity shifts of 
observed Ly$\alpha$ peaks produced by HI absorption systems as a function of 
absorber offset velocity. 
Plots are for different N(HI) as indicated (in cm$^{-2}$). Solid
lines are for $b=50$ km s$^{-1}$, dashed/dotted lines are lower and upper
values calculated for $b=25$ and 75 km s$^{-1}$. {\it Right)} Flux ratios of
Ly$\alpha$ before and after absorption with indicated HI column density. Solid
lines for $b=50$ km s$^{-1}$, lower and upper dashed/dotted lines for $b=75$
and 25 km s$^{-1}$. For more details see text
\end{figure}

\newpage \clearpage

\begin{figure}[p]
\centerline{
\psfig{figure=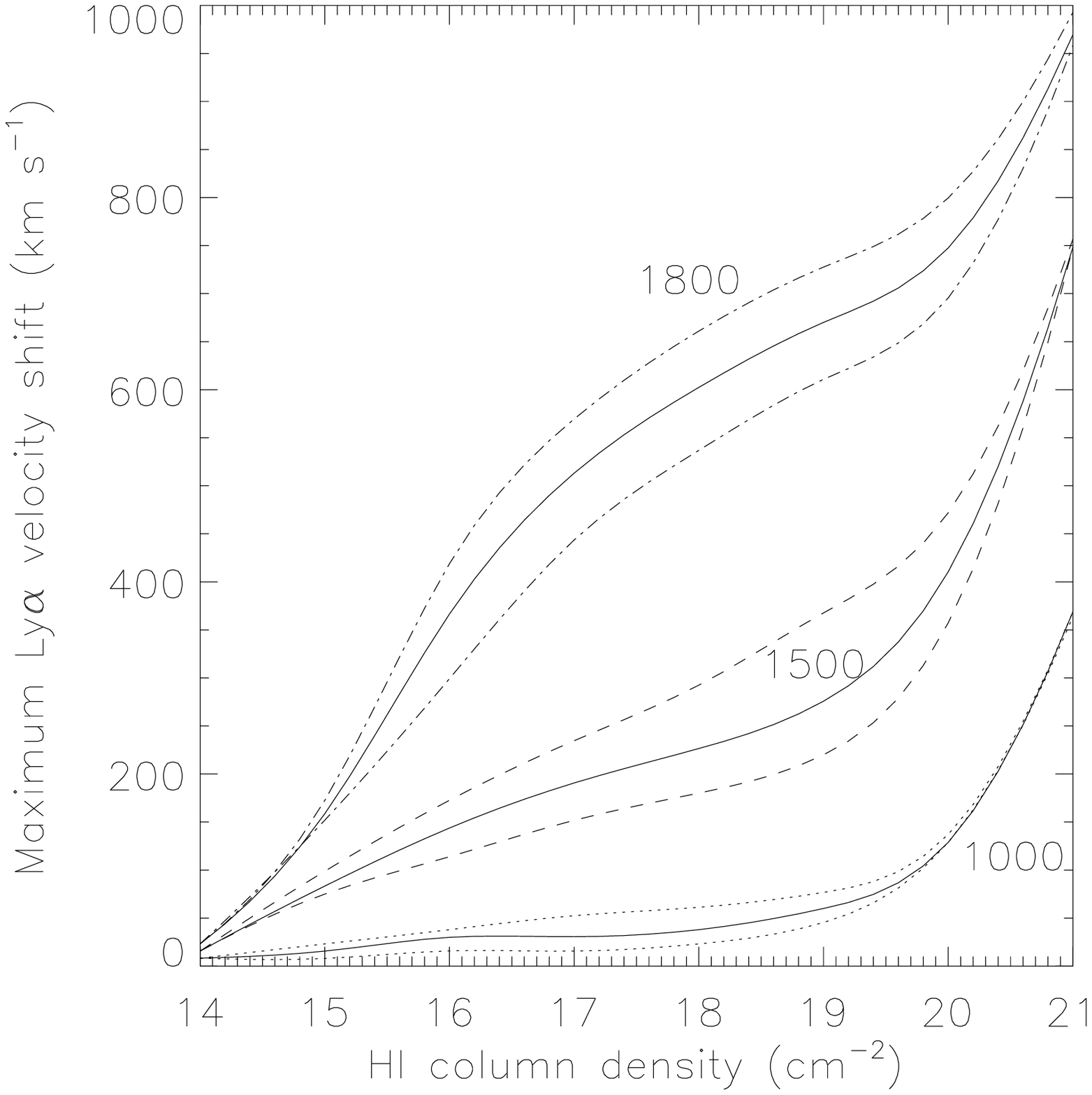,width=8cm}
}

\noindent{\bf Fig. 8.} Maximum velocity shifts of the peak of Ly$\alpha$ as a
function of HI absorption column density. The solid
lines are for the indicated original widths of Ly$\alpha$ emission 
(in units km s$^{-1}$ FWHM) and $b=50$ km s$^{-1}$.
Lower and upper dashed/dotted lines correspond to $b=25$ and $b=75$ km s$^{-1}$
\end{figure}

\newpage \clearpage

\begin{figure}[p]
\hbox{
\psfig{figure=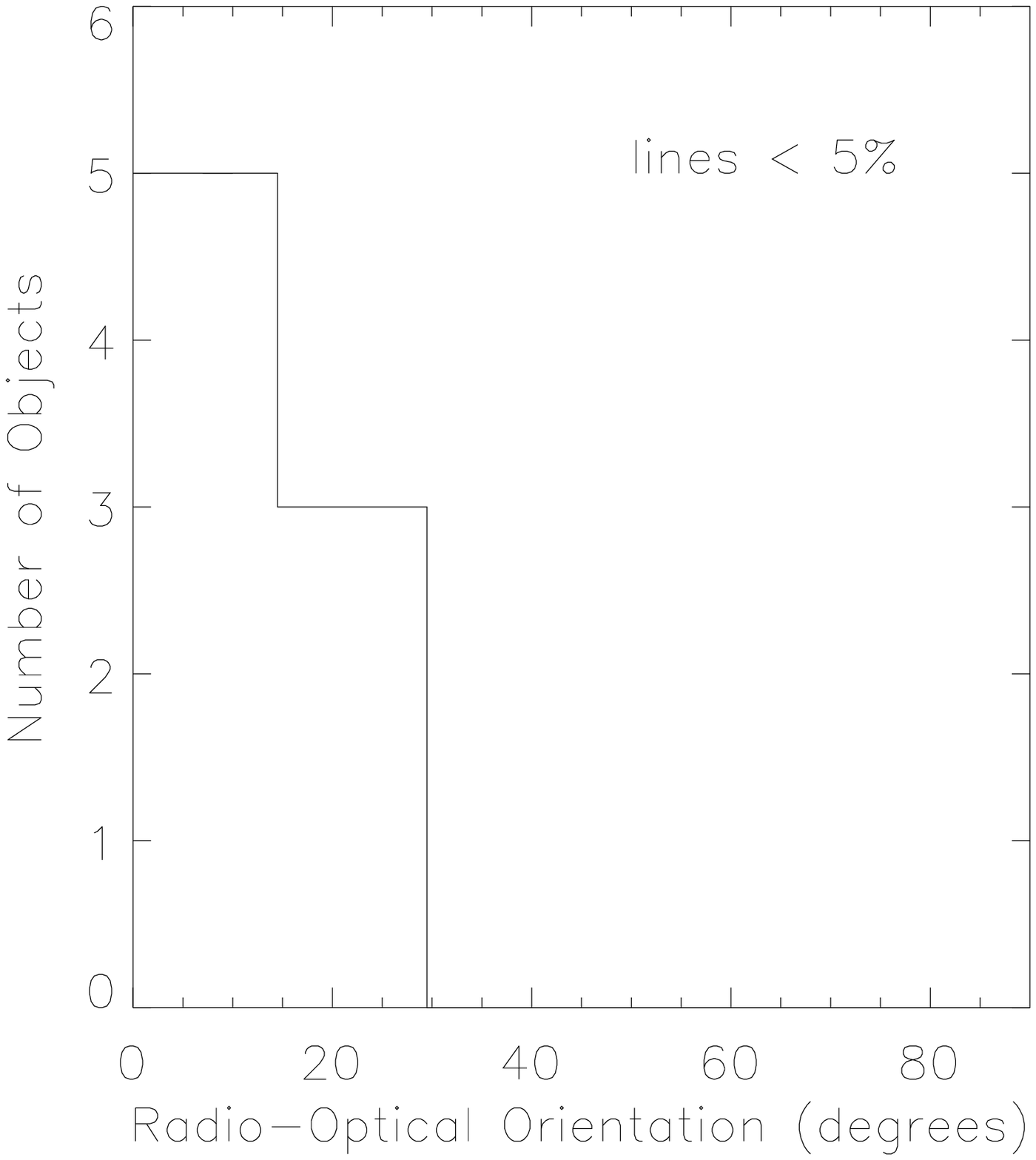,width=8cm}
\psfig{figure=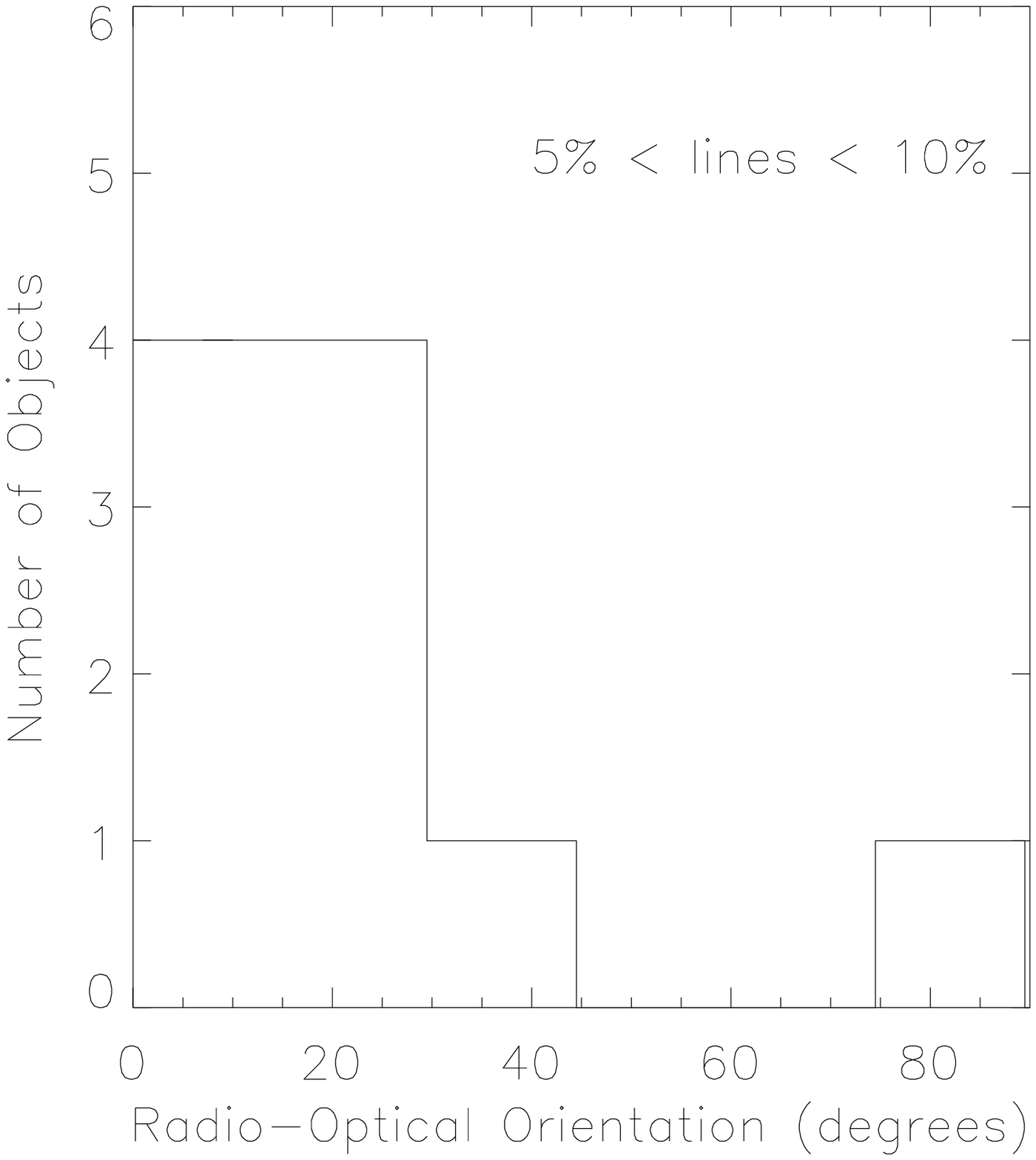,width=8cm}
}

\noindent{\bf Fig. 9.} Difference between the orientation of the 
optical $R$ band and the radio axes for 18 USS sources with $1.9<z<3$.
At these redshifts there are no strong emission lines with wavelength
within the filter bandpass. The alignment is similarly strong for
objects with line contaminations in $R$ band of 5\%--10\% as for those 
objects with less than 5\% line contamination
\end{figure}

\newpage \clearpage 

\begin{figure}[p]
\centerline{
\psfig{figure=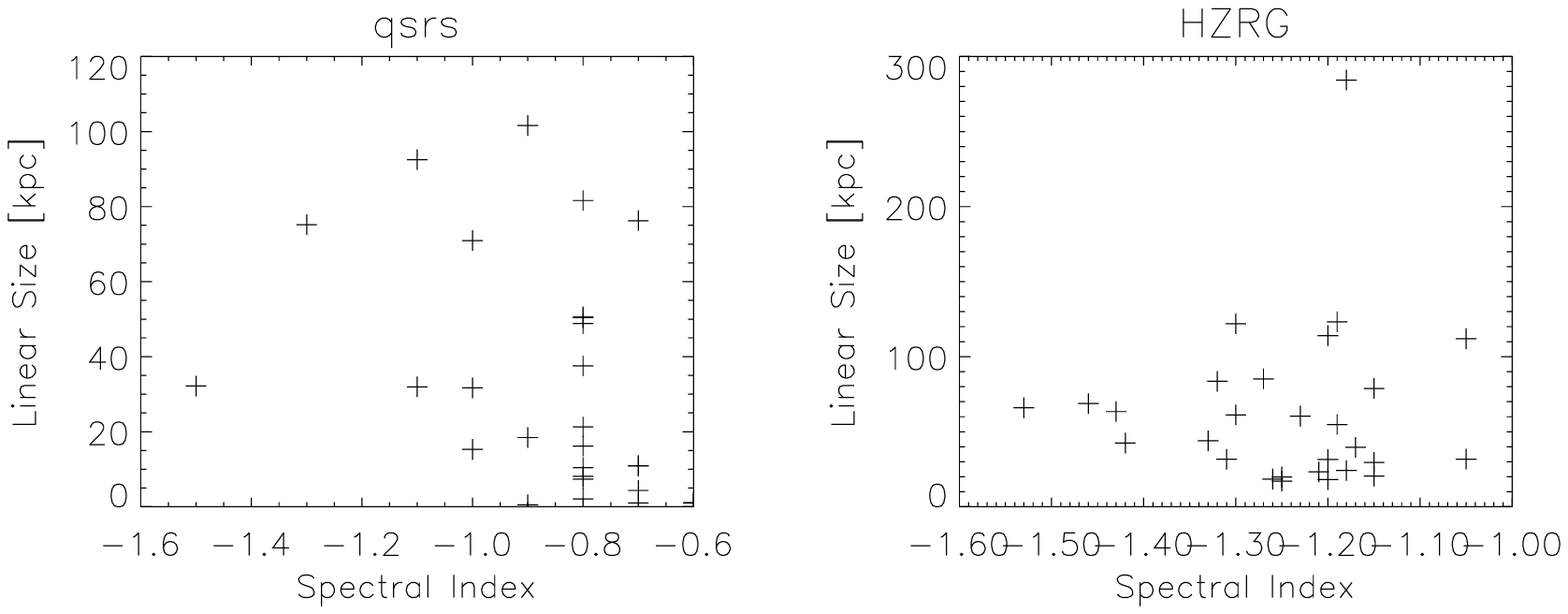}}

\noindent{\bf Fig. 10.} Comparison of the projected linear sizes for $z>2$ 
radio sources. Left shows the steep spectrum
quasars from the sample of Barthel and Miley. Right shows the 
distribution for our USS radio galaxies
\end{figure}

\newpage \clearpage 

\begin{figure}[p]
\centerline{
\psfig{figure=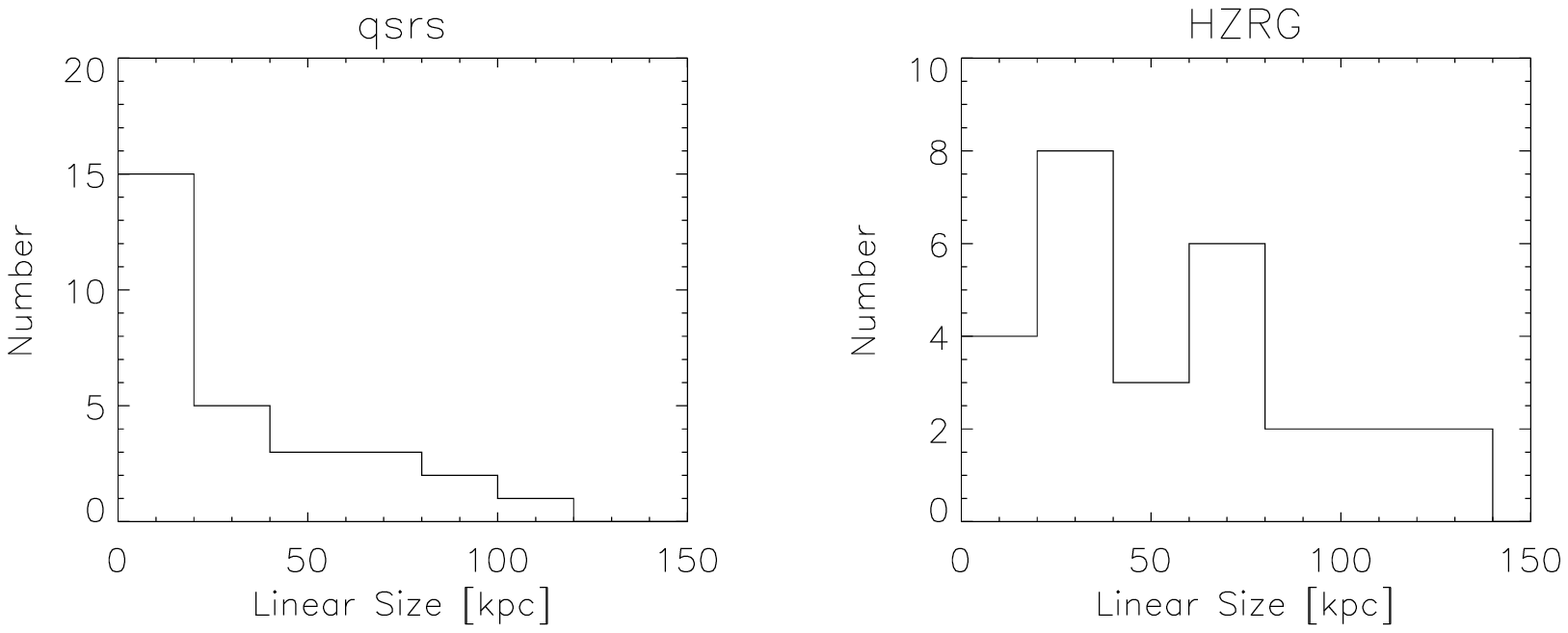}
}

\noindent{\bf Fig. 11.} The linear size as a function of spectral index for 
the $z>2$ USS radio galaxies (left) 
and quasars from the Barthel and Miley sample (right)
\end{figure}

\newpage \clearpage 
\section{Tables}

\vspace{2cm}

{\small
\begin{table}
\vspace{-0.5cm}
{\bf Table 1.} Observational setup
\begin{center}
\begin{tabular}{@{\extracolsep{-0.1em}}l|ccccc} 
\hline 
Dates                               &Sep 90        &Nov 90            &Mar 91            &Nov 91           &Jan 92                \\
                                    &              &May 91            &                  &Apr 92           &                      \\  
                                    &              &                  &                  &Nov 92           &                      \\  
\hline  			                                  												    
Telescope                           &NTT           &ESO 3.6           &NTT               &ESO 3.6          &WHT                   \\
Instrument                          &EFOSC2        &EFOSC             &EMMI              &EFOSC            &ISIS                  \\
CCD                                 &Thompson\# 17 &RCA \# 8          &Thompson          & TEK \# 26       &EEV 4                 \\
CCD size                            &$1024^2$      &$1024\times 640$  &$1100\times 1040$ &$512 \times 512$ &$1280 \times 1180 $   \\
Binning                             &$2 \times 4 $ &$2\times2$        &$-$               &$ 1 \times 2$    &$3 \times 2$          \\
Binned Pixel                        &0.7           &$0.67''$          &$0.44''$          &$0.61''$         &$0.74''$              \\
Slit                                &$2.2''$       &$2.5''$           &$2''$             &$2.5''$          &$2.5''$               \\
Grism(s)/gratings                   &UV 300 /      &B300              &Grism 3           &B300             &R158B                 \\
                                    &B300                                                                                   \\
Spectral ranges [$10^3 \times$\ang] &$3.6 - 5.4$ / &$3.6- 7$          &$4.1 - 7.9$       &$3.8 - 6.9$      & $3.6 - 6.9$          \\
                                    &$ 4.7 - 6.8 $                                                                            \\ 
Wavelength resolution               &25 \ang       &24 \ang           &18  \ang          &24 \ang          &20 \ang               \\
\hline 
\end{tabular}
\end{center}
\end{table} 
} 

{\small
\begin{table}[htp]
\vspace{-1cm}
{\bf Table 2.} Redshifts and emission line properties for $z>1.9$ objects
\begin{center}
\begin{tabular}{@{\extracolsep{-0.25em}}cr@{\,}rcr@{\,}rr@{\,}rr@{\,}rr@{\,}rc@{\,}rr}
\hline \\[-0.8em]
 Name&\multicolumn{2}{c}{z}&       Line&    \multicolumn{2}{c}{Peak}&   
\multicolumn{2}{c}{Flux}& \multicolumn{2}{c}{FWHM}& \multicolumn{2}{c}{Eq. width}&Ly$\alpha$\\ 
     &              &   & &            \multicolumn{2}{c}{Wavelength}   &  \multicolumn{2}{c}{$10^{-16}$} & 
             &        & & & Extent         \\
     &              &   &               & \multicolumn{2}{c}{(\AA)}&\multicolumn{2}{r}{(erg s$^{-1}$cm$^{-2}$)}&\multicolumn{2}{c}{(km s$^{-1}$)}&\multicolumn{2}{c}{(\ang)}  &($''$)\\
\\[-0.8em]
\hline 
\\[-0.8em]
0200$+$015& 2.229&$\pm$ 0.002&
   \la & 3927&$\pm$  2&\quad 17.4&$\pm$ 2.2&       $<  681$&               &       $>  256$&        &       6\\
 & & & 
   CIV & 5002&  3&\quad  4.2& 0.5&           1658&            624&             63&             13\\
 & & & 
  HeII & 5297&  2&\quad  3.2& 0.4&            890&            506&             36&              5\\
 & & & 
 CIII] & 6158&  2&\quad  4.0& 0.5&           1423&            487&             58&              9\\
\\[-0.8em]
0211$-$122& 2.336&$\pm$ 0.002&
   \la & 4053&$\pm$  1&\quad  5.7&$\pm$ 0.6&           1698&     $\pm$  635&             49&     $\pm$    6& 10\\
 & & & 
    NV & 4139&  2&\quad  4.1& 0.5&           1812&            666&             35&              4\\
 & & & 
SiIV/O & 4674&  6&\quad  0.6& 0.2&       $<  736$&               &              5&              2\\
 & & & 
   CIV & 5168&  1&\quad  5.6& 0.6&           1225&            472&             48&              5\\
 & & & 
  HeII & 5472&  1&\quad  3.1& 0.4&            898&            453&             33&              4\\
 & & & 
 CIII] & 6364&  1&\quad  2.2& 0.3&            610&            411&             27&              3\\
\\[-0.8em]
0214$+$183& 2.130&$\pm$ 0.003&
   CIV & 4852&$\pm$  1&\quad  3.0&$\pm$ 0.3&           1142&     $\pm$  516&             81&     $\pm$   12 & \\
 & & & 
  HeII & 5125&  4&\quad  1.8& 0.2&           2016&            586&             50&              8\\
 & & & 
 CIII] & 5971&  2&\quad  1.8& 0.2&            801&            423&             38&              5\\
\\[-0.8em]
0355$-$037& 2.153&$\pm$ 0.001&
   \la & 3836&$\pm$  2&\quad 11.2&$\pm$ 1.8&       $<  667$&               &       $>  263$&               & 8\\
 & & & 
   CIV & 4883&  4&\quad  2.7& 0.6&            859&            690&             26&              7\\
 & & & 
  HeII & 5170&  4&\quad  3.7& 0.7&           1514&            687&             62&             18\\
 & & & 
 CIII] & 6018&  3&\quad  2.3& 0.5&       $<  470$&               &             33&              9\\
\\[-0.8em]
0417$-$181& 2.773&$\pm$ 0.003&
   \la & 4594&$\pm$  2&\quad  3.0&$\pm$ 0.3&           2303&     $\pm$  590&             17&     $\pm$    1& 3 \\
 & & & 
   CIV & 5842&  3&\quad  1.2& 0.2&           1047&            510&              8&              1\\
 & & & 
  HeII & 6185&  2&\quad  0.5& 0.1&       $<  430$&               &              3&              0\\
\\[-0.8em]
0448$+$091& 2.037&$\pm$ 0.002&
   \la & 3692&$\pm$  1&\quad 12.2&$\pm$ 1.4&       $<  558$&               &       $>  321$&               & 13\\
 & & & 
   CIV & 4717& 10&\quad  1.2& 0.4&           1542&            885&       $>  101$&               \\
 & & & 
  HeII & 4986&  4&\quad  1.4& 0.4&            688&            562&       $>   40$&               \\
 & & & 
 CIII] & 5789&  7&\quad  2.7& 0.6&           1709&            930&       $>   60$&               \\
\\[-0.8em]
0529$-$549& 2.575&$\pm$ 0.002&
   \la & 4345&$\pm$  1&\quad  7.4&$\pm$ 0.8&           1944&     $\pm$  591&             39&     $\pm$    4& 5\\
 & & & 
   CIV & 5539&  3&\quad  0.4& 0.1&       $<  530$&               &              2&              0\\
 & & & 
  HeII & 5866&  3&\quad  0.6& 0.1&       $<  524$&               &              4&              0\\
 & & & 
 CIII] & 6833& 10&\quad  1.8& 0.3&           2322&           1047&             11&              1\\
\\[-0.8em]
0748$+$134& 2.419&$\pm$ 0.004&
   \la & 4161&$\pm$  1&\quad  6.3&$\pm$ 0.8&           1311&     $\pm$  503&       $>  522$&               & 5\\
 & & & 
   CIV & 5292&  3&\quad  1.8& 0.3&           1022&            526&             77&             29\\
 & & & 
  HeII & 5598&  5&\quad  1.5& 0.3&           1549&            513&       $>  209$&               \\
 & & & 
 CIII] & 6515&  6&\quad  1.4& 0.3&           1293&            671&       $>   62$&               \\
\\[-0.8em]
0828$+$193& 2.572&$\pm$ 0.002&
   \la & 4342&$\pm$  1&\quad 13.3&$\pm$ 1.3&           1166&     $\pm$  548&       $> 2730$&               & 7\\
 & & & 
   CIV & 5534&  1&\quad  1.9& 0.2&       $<  435$&               &             43&              5\\
 & & & 
  HeII & 5862&  2&\quad  1.9& 0.2&           1576&            445&             41&              5\\
 & & & 
 CIII] & 6813&  2&\quad  2.0& 0.2&           1261&            378&             39&              4\\
\\[-0.8em]
0857$+$036& 2.814&$\pm$ 0.003&
   \la & 4640&$\pm$  1&\quad  2.6&$\pm$ 0.3&           1095&     $\pm$  441&            113&     $\pm$   24& 2\\
 & & & 
   CIV & 5908&  3&\quad  1.0& 0.1&           1650&            426&             78&             17\\
 & & & 
  HeII & 6247&  2&\quad  0.7& 0.1&            956&            391&             47&             12\\

\\[-0.8em]
\hline
\end{tabular} 
\end{center}
\end{table}
}

{\small
\begin{table}[htp]
\begin{tabular}{@{\extracolsep{-0.25em}}cr@{\,}rcr@{\,}rr@{\,}rr@{\,}rr@{\,}rc@{\,}rr}
\hline \\[-0.8em]
 Name&\multicolumn{2}{c}{z}&       Line&    \multicolumn{2}{c}{Peak}&   
\multicolumn{2}{c}{Flux}& \multicolumn{2}{c}{FWHM}& \multicolumn{2}{c}{Eq. width}&Ly$\alpha$\\ 
     &              &   & &            \multicolumn{2}{c}{Wavelength}   &  \multicolumn{2}{c}{$10^{-16}$} & 
             &          & & &     Extent   \\
     &              &   &               & \multicolumn{2}{c}{(\AA)}&\multicolumn{2}{r}{(erg s$^{-1}$cm$^{-2}$)}&\multicolumn{2}{c}{(km s$^{-1}$)}&\multicolumn{2}{c}{(\ang)}  &($''$)\\
\\[-0.8em]
\hline 
\\[-0.8em]
0943$-$242& 2.923&$\pm$ 0.002&
   \la & 4772&$\pm$  0&\quad 20.1&$\pm$ 2.0&           1666&     $\pm$  264&       $>  653$&               & 3\\
 & & & 
   CIV & 6076&  2&\quad  3.9& 0.4&           1619&            348&            113&             37\\
 & & & 
  HeII & 6430&  1&\quad  2.7& 0.3&            874&            241&             83&             29\\
 & & & 
 CIII] & 7478&  8&\quad  2.3& 0.4&           1397&           1053&       $>   91$&               \\
\\[-0.8em]
1113$-$178& 2.239&$\pm$ 0.003&
   \la & 3953&$\pm$  6&\quad  6.4&$\pm$ 0.7&           5021&     $\pm$  764&             51&     $\pm$    3& \\
 & & & 
   CIV & 5042& 62&\quad  1.7& 0.4&          10196&           1530&              8&              1\\
 & & & 
  HeII & 5310&  3&\quad  0.7& 0.2&       $<  559$&               &              3&              0\\
 & & & 
 CIII] & 6180&  2&\quad  2.8& 0.3&           1708&            459&             16&              1\\
\\[-0.8em]
1138$-$262& 2.156&$\pm$ 0.004&
   \la & 3837&$\pm$  2&\quad 13.9&$\pm$ 1.6&           2603&     $\pm$  695&       $>  832$&               & \\
 & & & 
   CIV & 4900&  6&\quad  0.8& 0.2&           1982&            728&              6&              1\\
 & & & 
  HeII & 5166&  6&\quad  1.3& 0.2&           2787&           1005&             10&              1\\
 & & & 
 CIII] & 6017& 24&\quad  1.3& 0.3&           4285&           1341&             12&              2\\
\\[-0.8em]
1243$+$036& 3.570&$\pm$ 0.001&
   \la & 5556&$\pm$  1&\quad 23.5&$\pm$ 2.4&           2454&     $\pm$  251&             351&     $\pm$   13 & 7\\
\\[-0.8em]
1357$+$007& 2.673&$\pm$ 0.002&
   \la & 4467&$\pm$  1&\quad  9.1&$\pm$ 1.0&           2188&     $\pm$  571&             45&     $\pm$    4 & 3\\
 & & & 
   CIV & 5682&  5&\quad  1.7& 0.3&            957&            653&             13&              2\\
\\[-0.8em]
1410$-$001& 2.363&$\pm$ 0.001&
   \la & 4089&$\pm$  1&\quad 32.0&$\pm$ 3.2&           1266&     $\pm$  485&            356&     $\pm$   57 & 10\\
 & & & 
   CIV & 5209&  1&\quad  5.2& 0.6&           1330&            402&             78&             11\\
 & & & 
  HeII & 5516&  2&\quad  3.6& 0.4&           1617&            430&             48&              7\\
 & & & 
 CIII] & 6413&  3&\quad  3.3& 0.4&           1838&            444&             73&              13\\
\\[-0.8em]
1436$+$157& 2.538&$\pm$ 0.003&
   \la & 4302&$\pm$  3&\quad 42.0&$\pm$ 4.7&           3731&     $\pm$  733&             48&     $\pm$    5& 8\\
 & & & 
   CIV & 5474& 44&\quad 17.0& 1.9&           9174&           1397&              19&              2\\
 & & & 
  HeII & 5799&  9&\quad  6.0& 0.8&           2952&            998&              8&              1\\
 & & & 
 CIII] & 6730& 42&\quad  9.4& 1.7&           4531&           1956&              16&              2\\
\\[-0.8em]
1545$-$234& 2.755&$\pm$ 0.002&
   \la & 4567&$\pm$  1&\quad  6.7&$\pm$ 0.7&           1377&     $\pm$  534&       $>  900$&               & 5\\
 & & & 
   CIV & 5810&  4&\quad  1.1& 0.2&           1221&            562&             94&             40\\
 & & & 
  HeII & 6158&  4&\quad  1.8& 0.3&           1729&            560&       $>   86$&               \\
\\[-0.8em]
1558$-$003& 2.527&$\pm$ 0.002&
   \la & 4287&$\pm$  1&\quad 14.9&$\pm$ 1.5&           1529&     $\pm$  556&           1180&     $\pm$  510 & 6\\
 & & & 
   CIV & 5468&  2&\quad  2.7& 0.3&           1965&            501&             95&             15\\
 & & & 
  HeII & 5785&  5&\quad  1.7& 0.2&           2468&            613&             82&             17\\
 & & & 
 CIII] & 6739&  4&\quad  1.2& 0.2&           1389&            525&             73&             21\\
\\[-0.8em]
1707$+$105& 2.349&$\pm$ 0.005&
   \la & 4076&$\pm$  1&\quad  4.3&$\pm$ 0.5&           1466&     $\pm$  501&            242&     $\pm$   80& 16\\
 & & & 
  HeII & 5480&  3&\quad  0.9& 0.2&            972&            475&             70&             24\\
 & & & 
 CIII] & 6384& 22&\quad  0.2& 0.1&           1692&            940&       $>    7$&               \\
\\[-0.8em]
2202$+$128& 2.706&$\pm$ 0.001&
   \la & 4506&$\pm$  1&\quad  7.7&$\pm$ 0.8&           1167&     $\pm$  532&       $> 1375$&               & 3\\
 & & & 
   CIV & 5736&  5&\quad  1.9& 0.4&           1464&            693&       $>  225$&               \\
 & & & 
  HeII & 6064& 14&\quad  2.4& 0.5&           3023&           2071&       $>  309$&               \\
\\[-0.8em]
2251$-$089& 1.986&$\pm$ 0.002&
   CIV & 4628&$\pm$  2&\quad  3.3&$\pm$ 0.4&            985&     $\pm$  578&       $>  148$&               \\
 & & & 
  HeII & 4895&  2&\quad  1.3& 0.2&       $<  549$&               &             66&             19\\
 & & & 
 CIII] & 5696&  2&\quad  1.5& 0.2&           1016&            498&             84&             20\\

\\[-0.8em]
\hline
\end{tabular} 
\end{table}
}

{\small
\begin{table}[htp]
\begin{tabular}{@{\extracolsep{-0.25em}}cr@{\,}rcr@{\,}rr@{\,}rr@{\,}rr@{\,}rc@{\,}rr}
\hline \\[-0.8em]
 Name&\multicolumn{2}{c}{z}&       Line&    \multicolumn{2}{c}{Peak}&   
\multicolumn{2}{c}{Flux}& \multicolumn{2}{c}{FWHM}& \multicolumn{2}{c}{Eq. width}&Ly$\alpha$\\ 
     &              &   & &            \multicolumn{2}{c}{Wavelength}   &  \multicolumn{2}{c}{$10^{-16}$} & 
             &             & & &    Extent \\
     &              &   &               & \multicolumn{2}{c}{(\AA)}&\multicolumn{2}{r}{(erg s$^{-1}$cm$^{-2}$)}&\multicolumn{2}{c}{(km s$^{-1}$)}&\multicolumn{2}{c}{(\ang)}  &($''$)\\
\\[-0.8em]
\hline 
\\[-0.8em]
 4C23$.$56& 2.483&$\pm$ 0.003&
   CIV & 5388&$\pm$ 10&\quad  5.1&$\pm$ 0.6&           3695&     $\pm$ 1020&            149&     $\pm$   38\\
 & & & 
  HeII & 5700& 10&\quad  1.3& 0.3&           1636&            896&             31&              8\\
 & & & 
 CIII] & 6651&  4&\quad  1.5& 0.3&            734&            626&             34&              8\\
\\[-0.8em]
 4C24$.$28& 2.879&$\pm$ 0.006&
   \la & 4714&$\pm$  3&\quad  7.3&$\pm$ 1.0&       $<  762$&               &             72&     $\pm$  12& 5\\
 & & & 
    NV & 4835& 12&\quad  6.9& 1.1&           3562&           2149&             70&             19\\
 & & & 
   CIV & 6028& 24&\quad  1.7& 0.4&           3284&           2610&             20&              6\\
\\[-0.8em]
 4C26$.$38& 2.608&$\pm$ 0.001&
   CIV & 5587&$\pm$  2&\quad  8.9&$\pm$ 1.1&       $<  560$&               &       $>  156$&               \\
 & & & 
  HeII & 5919&  4&\quad  5.7& 0.8&           1367&            734&             116&             46\\
 & & & 
 CIII] & 6888&  5&\quad  2.4& 0.5&       $<  603$&               &       $>   45$&               \\
\\[-0.8em]
 4C28$.$58& 2.891&$\pm$ 0.004&
   CIV & 6033&$\pm$ 10&\quad  0.3&$\pm$ 0.2&       $<  615$&               &       $>    1$&               \\
 & & & 
  HeII & 6399& 14&\quad  1.6& 0.4&           2186&           1007&             9&              2\\
 & & & 
 CIII] & 7424&  5&\quad  1.8& 0.4&            569&            520&             12&              2\\
\\[-0.8em]
 4C40$.$36& 2.265&$\pm$ 0.003&
   CIV & 5048&$\pm$  4&\quad  6.2&$\pm$ 0.8&           2017&     $\pm$  782&             54&     $\pm$    9\\
 & & & 
  HeII & 5356&  3&\quad  5.6& 0.7&           1640&            718&             54&              9\\
 & & & 
 CIII] & 6235&  5&\quad  5.9& 0.6&           2370&            668&             67&              10\\
\\[-0.8em]
 4C41$.$17& 3.792&$\pm$ 0.001&
   \la & 5827&$\pm$  1&\quad 22.0&$\pm$ 2.3&            597&     $\pm$  514&            159&     $\pm$   31& 11\\
 & & & 
   CIV & 7422& 14&\quad  1.0& 0.5&       $<  685$&               &       $>    4$&               \\
\\[-0.8em]
 4C48$.$48& 2.343&$\pm$ 0.003&
   CIV & 5183&$\pm$  3&\quad  6.1&$\pm$ 0.9&       $<  723$&               &             56&     $\pm$   13\\
 & & & 
  HeII & 5485&  4&\quad  3.7& 0.6&       $<  766$&               &             44&             11\\
 & & & 
 CIII] & 6376&  3&\quad  2.8& 0.4&       $<  531$&               &             26&              4\\
\\[-0.8em]
 4C60$.$07& 3.788&$\pm$ 0.004&
   \la & 5831&$\pm$  9&\quad 10.1&$\pm$ 1.3&           2875&     $\pm$  937&            153&     $\pm$   71& 7\\
 & & & 
   CIV & 7412&  6&\quad  2.7& 0.8&       $<  740$&               &       $>   90$&               \\
 
\\[-0.8em]
\hline
\end{tabular} 
\end{table}
}

\begin{table}[htp]
\vspace{-2cm}
{\bf Table 3.} Redshifts of objects from the Leiden USS compendium with $z<1.9$
\begin{center}
\begin{tabular}{llll}  \hline
  Name & $z$  & Session & Comments \\ \hline
  0008+172 &1.390 &nov90& \\
  0036+205 &1.370 &aug90&\\
  0050+204 &1.297 &aug90&\\
  0054+090 &1.301 &nov91&\\
  0203$-$478 &0.836 &nov93&\\
  0241+348 &1.215 &aug90&\\
  0255+114 &(0.447) &jan92&\\
  0256+324 &(1.657) &dec89&\\
  0324$-$228 &1.898 &jan91&\\
  0338$-$259 &0.440 &nov93&\\
  0410$-$198 &(0.793) &nov91&\\
  0429$-$267 &1.26  &nov91&\\
  0447$-$164 &1.814 &nov90&\\
  0548$-$150 &0.650 &nov93&\\
  0729$-$722 &0.738 &nov91&\\
  0819+672 &0.760 &jan91&\\
  0850+140 &1.106 &jan92 & Quasar \\
  1035+208 &0.696 &may91&\\
  1043$-$216 &1.060 &mar91&\\
  1104+089 &1.384 &jan92&\\
  1228$-$166 &1.127 &jan92 & Quasar \\
  1238$-$273 &1.253 &may91&\\
  1449$-$004 &1.466 &may91&\\
  1502+039 &1.652 &may91&\\
  1548$-$111 &0.959 &apr92&\\
  1635+396 &1.600 &jul91&\\
  1657+003 &1.110 &may91&\\
  1725+167 &1.508 &may91&\\
  2000$-$091 &1.238 &may91&\\
  2214+134 &0.374 &aug90&\\
  2224$-$273 &1.678 &sep90&\\
  2226+162 &1.519 &nov91&\\
  2245+181 &0.830 &aug90&\\
  2339+269 &0.882 &aug90&\\ \hline
\end{tabular}
\end{center}
\end{table}

\begin{table}
{\bf Table 4.} Measured velocity shifts between emission lines
\begin{center}
\begin{tabular}{@{\extracolsep{-0.5em}}lrrrrrr} \hline
          & \multicolumn{6}{c}  {Velocity shifts (km s$^{-1}$)} \\
name      &CIV-Ly$\alpha$&HeII-Ly$\alpha$&CIII]-Ly$\alpha$&CIV-HeII& CIV-CIII & HeII-CIII \\ \hline
0200+015  &$-185\pm 285$&$-83\pm220 $&$ 324\pm 226$&$-101\pm 257$&$-509\pm 263$&$-408\pm 190$ \\
0211$-$122&$-528\pm 156$&$-315\pm160$&$ -53\pm 156$&$-213\pm 131$&$-474\pm 126$&$-261\pm 132$ \\
0214+183  &             &            &             &$-867\pm 265$&$-623\pm 149$&$245\pm 263$ \\
0355$-$037&$-19\pm 325 $&$ 221\pm326$&$ 206\pm 233$&$-241\pm 386$&$-226\pm 312$&$14\pm 313$ \\
0417$-$181&$ 298\pm 218$&$ 532\pm185$&             &$-234\pm 190$&            &   \\
%0447     &             &            &             &$24\pm 189$  &$-249\pm 162$&$-274\pm 161$ \\
0748+134  &$ 295\pm 210$&$ 756\pm326$&$ 811\pm 315$&$-461\pm 350$&$-516\pm 340$&$-55\pm 421$ \\
0857+036  &$-80\pm 177 $&$ 503\pm165$&             &$-583\pm 207$&            &  \\
0943$-$242&$-62\pm 148 $&$ 279\pm96 $&$ 551\pm 361$&$-341\pm 159$&$-614\pm 382$&$-273\pm 365$ \\
1138$-$262&$-968\pm 433$&$ 465\pm448$&$317\pm 1216$&$-1429\pm 562$&$-1281\pm 1262$&$148\pm 1268$ \\
1357+007  &$ 242\pm 295$&            &             &             &            &  \\
1410$-$001&$-197\pm 115$&$-22\pm152 $&$ 330\pm 193$&$-174\pm 158$&$-527\pm 198$&$-353\pm 222$ \\
1545$-$234&$227\pm 237 $&$  91\pm250$&             &$135\pm 314$ &           &  \\
1558$-$003&$-599\pm 165$&$-194\pm276$&$-431\pm 231$&$-403\pm 297$&$-168\pm 257$&$236\pm 339$ \\
1707+105  &             &$ 984\pm195$&$ 734\pm1063$&             &            &$251\pm 1071$ \\
2202+128  &$ 41\pm 294 $&$ 690\pm737$&             &$-648\pm 783$&            &  \\
2251$-$089&             &            &             &$-482\pm 210$&$-582\pm 204$&$-100\pm 212$ \\
4C23.56   &             &            &             &$-443\pm 781$&$311\pm 612$&$754\pm 576$ \\
4C24.28   &$-1336\pm 1235$&          &             &             &            &  \\
4C28.58   &             &            &             &$362\pm 861$ &$-626\pm 563$&$-990\pm 733$ \\
4C40.36   &             &            &             &$456\pm 355$ &$457\pm 383$&$1\pm 333$ \\
4C41.17   &$-165\pm 573$&            &             &             &            &  \\
4C48.48   &             &            &             &$-302\pm 293$&$-743\pm 233 $&$-441\pm 271$ \\
4C60.07   &$ 464\pm 564$&            &             &             &            &  \\ \hline
\end{tabular}
\end{center}
\end{table}

\end{document}